\documentclass[preprint,aps,prd,USpaper,nofootinbib,showpacs,superscriptaddress]{revtex4-2}
\usepackage{amsmath,amssymb,graphicx,mathrsfs,xcolor}
\usepackage{bm} 
\usepackage{ulem}
\usepackage{feynmp}

\usepackage{ifpdf}
\ifpdf
\fi
\makeatletter
\def\endfmffile{%
  \fmfcmd{\p@rcent\space the end.^^J%
          end.^^J%
          endinput;}%
  \if@fmfio
    \immediate\closeout\@outfmf
  \fi
  \IfFileExists{\thefmffile.mp}{\immediate\write18{mpost \thefmffile}}{}
  \let\thefmffile\relax
}
\makeatother
\newcommand{\nn}{\nonumber\\}

\newcommand{\Vg}{V_{\mbox{\scriptsize gluon}} }
\newcommand{\Vc}{V_{\mbox{\scriptsize conf}}}

\newcommand{\bGamma}{\bar{\Gamma}}

\newcommand{\bPsi}{\bar{\Psi}}

\newcommand{\ben}{\begin{displaymath}}
\newcommand{\een}{\end{displaymath}}
\newcommand{\be}{\begin{equation}}
\newcommand{\ee}{\end{equation}}
\newcommand{\bea}{\begin{eqnarray}}
\newcommand{\eea}{\end{eqnarray}}

\newcommand{\A}{\alpha}

\newcommand{\Gf}{G^{(4)}}
\newcommand{\Kf}{K^{d}}

\newcommand{\q}{\bar{q}}
\newcommand{\D}{\bar{D}}

\newcommand{\PPhi}{{\it\Phi}}
\newcommand{\GG}{{\it\Gamma}}
\newcommand{\bGG}{{\it \bar\Gamma}}
\newcommand{\RR}{{\cal R}}
\newcommand{\PP}{{\cal P}}
\newcommand{\TT}{{\cal T}}

\newcommand{\bc}{\begin{center}}
\newcommand{\ec}{\end{center}}

\newcommand{\eqn}[1]{\label{#1}}
\newcommand{\eq}[1]{Eq.~(\ref{#1})}
\newcommand{\eqs}[1]{Eqs.~(\ref{#1})}
\newcommand{\fign}[1]{\label{#1}}
\newcommand{\fig}[1]{Fig.~\ref{#1}}

\begin{document}
\title{Unified tetraquark equations}
\author{A. N. Kvinikhidze}
\email{sasha\_kvinikhidze@hotmail.com}
\affiliation{Andrea Razmadze Mathematical Institute of Tbilisi State University, 6, Tamarashvili Str., 0186 Tbilisi, Georgia}
\affiliation{
College of Science and Engineering,
 Flinders University, Bedford Park, SA 5042, Australia}
\author{B. Blankleider}
\email{boris.blankleider@flinders.edu.au}
\affiliation{
College of Science and Engineering,
 Flinders University, Bedford Park, SA 5042, Australia}

\date{\today}

\begin{abstract}
We derive covariant equations describing the tetraquark in terms of an admixture of two-body states $D\bar D$ (diquark-antidiquark), $MM$ (meson-meson), and three-body-like states $q\bar q (T_{q\q})$, $q q (T_{\q\q})$, and $\q\q (T_{qq})$ where two of the quarks are spectators while the other two are interacting (their t matrices denoted correspondingly as $T_{q\q}$, $T_{\q\q}$, and $T_{qq}$). This has been achieved by describing the $qq\bar q\bar q$ system using the Faddeev-like four-body equations of Khvedelidze and Kvinikhidze [Theor.\ Math.\ Phys.\ {\bf 90}, 62 (1992)] while retaining all two-body interactions (in contrast to previous works where terms involving isolated two-quark scattering were neglected). As such, our formulation, is able to unify seemingly unrelated models of the tetraquark, like, for example, the $D\bar D$ model of the Moscow group [Faustov {\it et al.}, Universe {\bf 7}, 94 (2021)] and the coupled channel $D \bar D-MM$ model of the Giessen group [Heupel {\it et al.}, Phys.\ Lett.\ {\bf B718}, 545 (2012)].
\end{abstract}

\maketitle
\newpage

\section{Introduction}

With the inception of the quark model of hadrons in 1964, all known baryons and mesons could be described as stable combinations of valence quarks $q$ and antiquarks $\bar q$, baryons consisting of three quarks ($qqq$) and mesons of a quark-antiquark pair ($q\bar q$) \cite{Gell-Mann:1964ewy, Zweig:1964ruk}. Although multiquark states such as the tetraquark ($qq\bar q \bar q$) and pentaquark ($qqqq\bar q$) were also considered to be a possibility \cite{Gell-Mann:1964ewy,Jaffe_PRD15_267,*Jaffe_PRD15_281}, it wasn't until 2003 that the first experimental evidence for an exotic  multiquark state (a tetraquark) became available \cite{Belle:2003nnu}. Since then there has been a virtual explosion in the number of multiquark hadron candidates discovered, together with a correspondingly large variety of theoretical models developed in order to learn about the dynamics of their formation, see \cite{Chen:2022asf} for a recent review.

Out of the many recent theoretical works on this subject, we would like to address the works of the Moscow group (Faustov {\it et al.})\! \cite{Ebert:2005nc,Faustov:2020qfm,Faustov:2021hjs,Faustov:2022mvs}, who modeled tetraquarks as a diquark-antidiquark  $(D\bar D)$ system, and the Giessen group (Fischer  {\it et al.})\! \cite{Heupel:2012ua,Eichmann:2015cra,Eichmann:2020oqt, Santowsky:2021bhy}, who modeled tetraquarks as a coupled mix of meson-meson $(MM)$ and diquark-antidiquark $(D\bar D)$ states. It has been noted that these works differ significantly not only in their prediction of heavy tetraquark masses \cite{Faustov:2021hjs}, but moreover, in the very attribution of the inner structure a heavy tetraquark, with the Giessen group finding the $MM$ components to be generally dominant, with the $D\bar D$ components being small or even negligible \cite{Santowsky:2021bhy}. In view of the strongly differing predictions made by these models, it would be interesting and important to express these seemingly unrelated models in terms of a common theoretical foundation. It is to this end that we have derived a universal set of tetraquark equations which produce both the above approaches in different approximations.

In order to demonstrate how a unified theoretical approach is achieved, we first note that the Moscow group's model can be viewed as being based on the solutions of the bound-state equation for the $D\bar D$-tetraquark amplitude $\phi_D$, as illustrated in \fig{MBSE}. As seen from this figure, the kernel of the equation consists of a single term where a $q\q$ pair scatters elastically in the presence of spectating $q$ and $\q$ quarks. More specifically, the Moscow model corresponds to the case where $T_{q\q}$, the t matrix describing the mentioned $q\q$ scattering, is expressed as a sum of two potentials
\be
T_{q\q} = V_{\mbox{\scriptsize gluon}} + V_{\mbox{\scriptsize conf}}   \eqn{t_moscow}
\ee
where $\Vg$ is the $q\q$ one-gluon-exchange potential and $\Vc$ is a local confining  potential.\footnote{ 
To be precise, the Moscow group uses quasipotential bound state form factors instead of the $D\rightarrow qq$ form factor $\Gamma_{12}(p,P)$ and the $\bar D \rightarrow \q\q$ form factor $\Gamma_{34}(p,P)$, appearing as small blue circles in \fig{MBSE}. Formally, this is equivalent to assuming that $\Gamma_{12}(p,P)$ and $\Gamma_{34}(p,P)$  do not depend on the longitudinal projection of the relative 4-momentum $p$ with respect to the total momentum $P$ of the two quarks or two antiquarks.} However, in this paper we shall consider the general case of the $T_{q\q}$ t matrix, and correspondingly refer to the intermediate state of the kernel of \fig{MBSE} as $q\q(T_{q\q})$.
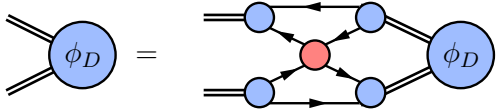
\begin{figure}[t]
\begin{center}
\begin{fmffile}{faust}
\[
\parbox{10mm}{
\begin{fmfgraph*}(10,15)
\fmfstraight
\fmfleftn{l}{7}\fmfrightn{r}{7}\fmfbottomn{b}{6}\fmftopn{t}{6}
\fmf{phantom}{l2,mb2,r2}
\fmf{phantom}{l6,mt2,r6}
\fmffreeze
\fmf{phantom}{r2,phi,r6}
\fmffreeze
\fmf{dbl_plain}{phi,l2}
\fmf{dbl_plain}{phi,l6}
\fmfv{d.sh=circle,d.f=empty,d.si=25,label=$\hspace{-4.5mm}\phi_D$,background=(.6235,,.7412,,1)}{phi}
\end{fmfgraph*}}
\hspace{6mm} = \hspace{5mm}
\parbox{37mm}{
\begin{fmfgraph*}(37,15)
\fmfstraight
\fmfleftn{l}{7}\fmfrightn{r}{7}\fmfbottomn{b}{7}\fmftopn{t}{7}
\fmf{phantom}{l2,vb1,mb1,vb2,mb2,r2}
\fmf{phantom}{l6,vt1,mt1,vt2,mt2,r6}
\fmf{phantom}{l2,xb1,mb1,xb2,mb2,r2}
\fmf{phantom}{l6,xt1,mt1,xt2,mt2,r6}
\fmf{phantom}{l2,yb1,mb1,yb2,mb2,r2}
\fmf{phantom}{l6,yt1,mt1,yt2,mt2,r6}
\fmf{phantom}{l4,l4a,l4b,r4b,r4a,r4}
\fmffreeze
\fmf{phantom}{l4,c,r4a}
\fmf{phantom}{l4,clb,r4a}
\fmf{phantom}{l4,clt,r4a}
\fmf{phantom}{l4,crb,r4a}
\fmf{phantom}{l4,crt,r4a}
\fmffreeze
\fmf{phantom}{r2,phi,r6}
\fmffreeze
\fmfshift{8 left}{phi}
\fmfshift{4 right}{xb1}
\fmfshift{4 down}{xb1}
\fmfshift{4 left}{xt2}
\fmfshift{4 down}{xt2}
\fmfshift{4 right}{xt1}
\fmfshift{4 down}{xt1}
\fmfshift{4 left}{xb2}
\fmfshift{4 down}{xb2}
\fmfshift{4 right}{yb1}
\fmfshift{4 up}{yb1}
\fmfshift{4 left}{yt2}
\fmfshift{4 up}{yt2}
\fmfshift{4 right}{yt1}
\fmfshift{4 up}{yt1}
\fmfshift{4 left}{yb2}
\fmfshift{4 up}{yb2}
\fmfshift{4 left}{clb}
\fmfshift{4 down}{clb}
\fmfshift{4 left}{clt}
\fmfshift{4 up}{clt}
\fmfshift{4 right}{crb}
\fmfshift{4 down}{crb}
\fmfshift{4 right}{crt}
\fmfshift{4 up}{crt}
\fmfv{d.sh=circle,d.f=empty,d.si=11,background=(.6235,,.7412,,1)}{vb1}
\fmfv{d.sh=circle,d.f=empty,d.si=11,background=(.6235,,.7412,,1)}{vb2}
\fmfv{d.sh=circle,d.f=empty,d.si=11,background=(.6235,,.7412,,1)}{vt1}
\fmfv{d.sh=circle,d.f=empty,d.si=11,background=(.6235,,.7412,,1)}{vt2}
\fmfv{d.sh=circle,d.f=empty,d.si=11,background=(1,,.51,,.5)}{c}
\fmf{dbl_plain}{l2,vb1}
\fmf{dbl_plain}{l6,vt1}
\fmf{dbl_plain}{phi,xb2}
\fmf{dbl_plain}{phi,yt2}
\fmfset{arrow_len}{2.5mm}
\fmfi{fermion}{vloc(__xt2) .. vloc(__crt)}
\fmfi{fermion}{ vloc(__clt)..vloc(__xt1)}
\fmfi{fermion}{vloc(__crb) ..vloc(__yb2)}
\fmfi{fermion}{vloc(__yb1) ..vloc(__clb)}
\fmfi{fermion}{vloc(__yt2) .. vloc(__yt1)}
\fmfi{fermion}{vloc(__xb1) .. vloc(__xb2)}
\fmfv{d.sh=circle,d.f=empty,d.si=25,label=$\hspace{-4.5mm}\phi_D$,background=(.6235,,.7412,,1)}{phi}
\end{fmfgraph*}}
\]
\end{fmffile}   
\vspace{-3mm}

\caption{\fign{MBSE}  Diquark-antidiquark bound state equation encompassing the Moscow group's approach \cite{Ebert:2005nc,Faustov:2020qfm,Faustov:2021hjs,Faustov:2022mvs}.  The form factor $\phi_D$ couples the tetraquark to diquark and antidiquark states (both represented by double-lines). Shown is the general form of the kernel where one $q \bar q$ pair interacts (the red circle representing the corresponding t matrix $T_{q\q}$) while the other $q \bar q$ pair is spectating. Quarks (antiquarks) are represented by left (right) directed lines.}
\end{center}
\end{figure}

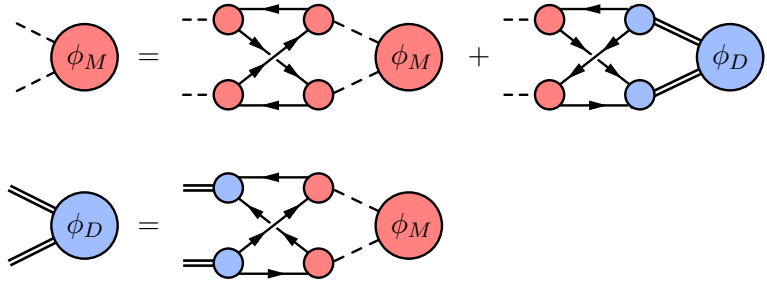
\begin{figure}[b]
\begin{center}
\begin{fmffile}{no2q}
\begin{align*}
\parbox{10mm}{
\begin{fmfgraph*}(10,15)
\fmfstraight
\fmfleftn{l}{7}\fmfrightn{r}{7}\fmfbottomn{b}{6}\fmftopn{t}{6}
\fmf{phantom}{l2,mb2,r2}
\fmf{phantom}{l6,mt2,r6}
\fmffreeze
\fmf{phantom}{r2,phi,r6}
\fmffreeze
\fmf{dashes}{phi,l2}
\fmf{dashes}{phi,l6}
\fmfv{d.sh=circle,d.f=empty,d.si=25,label=$\hspace{-4.5mm}\phi_M$,background=(1,,.51,,.5)}{phi}
\end{fmfgraph*}}
\hspace{6mm} &= \hspace{2mm}
\parbox{30mm}{
\begin{fmfgraph*}(30,15)
\fmfstraight
\fmfleftn{l}{7}\fmfrightn{r}{7}\fmfbottomn{b}{6}\fmftopn{t}{6}
\fmf{phantom}{l2,vb1,mb1,vb2,mb2,r2}
\fmf{phantom}{l6,vt1,mt1,vt2,mt2,r6}
\fmf{phantom}{l2,xb1,mb1,xb2,mb2,r2}
\fmf{phantom}{l6,xt1,mt1,xt2,mt2,r6}
\fmf{phantom}{l2,yb1,mb1,yb2,mb2,r2}
\fmf{phantom}{l6,yt1,mt1,yt2,mt2,r6}
\fmffreeze
\fmf{phantom}{r2,phi,r6}
\fmffreeze
\fmfshift{4 right}{xb1}
\fmfshift{4 down}{xb1}
\fmfshift{4 left}{xt2}
\fmfshift{4 down}{xt2}
\fmfshift{4 right}{xt1}
\fmfshift{4 down}{xt1}
\fmfshift{4 left}{xb2}
\fmfshift{4 down}{xb2}
\fmfshift{4 right}{yb1}
\fmfshift{4 up}{yb1}
\fmfshift{4 left}{yt2}
\fmfshift{4 up}{yt2}
\fmfshift{4 right}{yt1}
\fmfshift{4 up}{yt1}
\fmfshift{4 left}{yb2}
\fmfshift{4 up}{yb2}
\fmfv{d.sh=circle,d.f=empty,d.si=11,background=(1,,.51,,.5)}{vb1}
\fmfv{d.sh=circle,d.f=empty,d.si=11,background=(1,,.51,,.5)}{vb2}
\fmfv{d.sh=circle,d.f=empty,d.si=11,background=(1,,.51,,.5)}{vt1}
\fmfv{d.sh=circle,d.f=empty,d.si=11,background=(1,,.51,,.5)}{vt2}
\fmf{dashes}{l2,vb1}
\fmf{dashes}{l6,vt1}
\fmf{dashes}{phi,xb2}
\fmf{dashes}{phi,yt2}
\fmf{phantom}{vb1,vb2}
\fmf{phantom}{vt1,vt2}
\fmfset{arrow_len}{2.5mm}
\fmf{plain,rubout=3}{yb1,xt2}
\fmf{plain}{xt1,yb2}
\fmf{phantom_arrow}{xt1,c}
\fmf{phantom_arrow}{c,yb2}
\fmf{phantom_arrow,rubout=2}{yb1,c}
\fmf{phantom_arrow,rubout=2}{c,xt2}
\fmf{fermion}{yt2,yt1}
\fmf{fermion}{xb2,xb1}
\fmfv{d.sh=circle,d.f=empty,d.si=25,label=$\hspace{-4.5mm}\phi_M$,background=(1,,.51,,.5)}{phi}
\end{fmfgraph*}}
\hspace{7mm}+\hspace{1mm}
\parbox{30mm}{
\begin{fmfgraph*}(30,15)
\fmfstraight
\fmfleftn{l}{7}\fmfrightn{r}{7}\fmfbottomn{b}{6}\fmftopn{t}{6}
\fmf{phantom}{l2,vb1,mb1,vb2,mb2,r2}
\fmf{phantom}{l6,vt1,mt1,vt2,mt2,r6}
\fmf{phantom}{l2,xb1,mb1,xb2,mb2,r2}
\fmf{phantom}{l6,xt1,mt1,xt2,mt2,r6}
\fmf{phantom}{l2,yb1,mb1,yb2,mb2,r2}
\fmf{phantom}{l6,yt1,mt1,yt2,mt2,r6}
\fmffreeze
\fmf{phantom}{r2,phi,r6}
\fmffreeze
\fmfshift{4 right}{xb1}
\fmfshift{4 down}{xb1}
\fmfshift{4 left}{xt2}
\fmfshift{4 down}{xt2}
\fmfshift{4 right}{xt1}
\fmfshift{4 down}{xt1}
\fmfshift{4 left}{xb2}
\fmfshift{4 down}{xb2}
\fmfshift{4 right}{yb1}
\fmfshift{4 up}{yb1}
\fmfshift{4 left}{yt2}
\fmfshift{4 up}{yt2}
\fmfshift{4 right}{yt1}
\fmfshift{4 up}{yt1}
\fmfshift{4 left}{yb2}
\fmfshift{4 up}{yb2}
\fmfv{d.sh=circle,d.f=empty,d.si=11,background=(1,,.51,,.5)}{vb1}
\fmfv{d.sh=circle,d.f=empty,d.si=11,background=(.6235,,.7412,,1)}{vb2}
\fmfv{d.sh=circle,d.f=empty,d.si=11,background=(1,,.51,,.5)}{vt1}
\fmfv{d.sh=circle,d.f=empty,d.si=11,background=(.6235,,.7412,,1)}{vt2}
\fmf{dashes}{l2,vb1}
\fmf{dashes}{l6,vt1}
\fmf{dbl_plain}{phi,xb2}
\fmf{dbl_plain}{phi,yt2}
\fmf{phantom}{vb1,vb2}
\fmf{phantom}{vt1,vt2}
\fmfset{arrow_len}{2.5mm}
\fmf{plain,rubout=3}{xt2,yb1}
\fmf{plain}{yb2,xt1}
\fmf{phantom_arrow,rubout=2}{xt2,c}
\fmf{phantom_arrow,rubout=2}{c,yb1}
\fmf{phantom_arrow}{xt1,c}
\fmf{phantom_arrow}{c,yb2}
\fmf{fermion}{yt2,yt1}
\fmf{fermion}{xb1,xb2}
\fmfv{d.sh=circle,d.f=empty,d.si=25,label=$\hspace{-4.5mm}\phi_D$,background=(.6235,,.7412,,1)}{phi}
\end{fmfgraph*}}\\[6mm]
\parbox{10mm}{
\begin{fmfgraph*}(10,15)
\fmfstraight
\fmfleftn{l}{7}\fmfrightn{r}{7}\fmfbottomn{b}{6}\fmftopn{t}{6}
\fmf{phantom}{l2,mb2,r2}
\fmf{phantom}{l6,mt2,r6}
\fmffreeze
\fmf{phantom}{r2,phi,r6}
\fmffreeze
\fmf{dbl_plain}{phi,l2}
\fmf{dbl_plain}{phi,l6}
\fmfv{d.sh=circle,d.f=empty,d.si=25,label=$\hspace{-4.5mm}\phi_D$,background=(.6235,,.7412,,1)}{phi}
\end{fmfgraph*}}
\hspace{6mm} &= \hspace{2mm}
\parbox{30mm}{
\begin{fmfgraph*}(30,15)
\fmfstraight
\fmfleftn{l}{7}\fmfrightn{r}{7}\fmfbottomn{b}{6}\fmftopn{t}{6}
\fmf{phantom}{l2,vb1,mb1,vb2,mb2,r2}
\fmf{phantom}{l6,vt1,mt1,vt2,mt2,r6}
\fmf{phantom}{l2,xb1,mb1,xb2,mb2,r2}
\fmf{phantom}{l6,xt1,mt1,xt2,mt2,r6}
\fmf{phantom}{l2,yb1,mb1,yb2,mb2,r2}
\fmf{phantom}{l6,yt1,mt1,yt2,mt2,r6}
\fmffreeze
\fmf{phantom}{r2,phi,r6}
\fmffreeze
\fmfshift{4 right}{xb1}
\fmfshift{4 down}{xb1}
\fmfshift{4 left}{xt2}
\fmfshift{4 down}{xt2}
\fmfshift{4 right}{xt1}
\fmfshift{4 down}{xt1}
\fmfshift{4 left}{xb2}
\fmfshift{4 down}{xb2}
\fmfshift{4 right}{yb1}
\fmfshift{4 up}{yb1}
\fmfshift{4 left}{yt2}
\fmfshift{4 up}{yt2}
\fmfshift{4 right}{yt1}
\fmfshift{4 up}{yt1}
\fmfshift{4 left}{yb2}
\fmfshift{4 up}{yb2}
\fmfv{d.sh=circle,d.f=empty,d.si=11,background=(.6235,,.7412,,1)}{vb1}
\fmfv{d.sh=circle,d.f=empty,d.si=11,background=(1,,.51,,.5)}{vb2}
\fmfv{d.sh=circle,d.f=empty,d.si=11,background=(.6235,,.7412,,1)}{vt1}
\fmfv{d.sh=circle,d.f=empty,d.si=11,background=(1,,.51,,.5)}{vt2}
\fmf{dbl_plain}{l2,vb1}
\fmf{dbl_plain}{l6,vt1}
\fmf{dashes}{phi,xb2}
\fmf{dashes}{phi,yt2}
\fmf{phantom}{vb1,vb2}
\fmf{phantom}{vt1,vt2}
\fmfset{arrow_len}{2.5mm}
\fmf{plain,rubout=3}{yb1,xt2}
\fmf{plain}{yb2,xt1}
\fmf{phantom_arrow,rubout=2}{yb1,c}
\fmf{phantom_arrow,rubout=2}{c,xt2}
\fmf{phantom_arrow}{yb2,c}
\fmf{phantom_arrow}{c,xt1}
\fmf{fermion}{yt2,yt1}
\fmf{fermion}{xb1,xb2}
\fmfv{d.sh=circle,d.f=empty,d.si=25,label=$\hspace{-4.5mm}\phi_M$,background=(1,,.51,,.5)}{phi}
\end{fmfgraph*}}
\end{align*}
\end{fmffile}   
\vspace{-3mm}

\caption{\fign{GBSE}  Tetraquark equations of the Giessen group \cite{Heupel:2012ua,Eichmann:2015cra,Eichmann:2020oqt, Santowsky:2021bhy}. Form factor $\phi_M$ couples the tetraquark to two mesons (dashed lines), and form factors $\phi_D$ couples the tetraquark to diquark-antidiquark states (double-lines).}
\end{center}
\end{figure}
In a similar way, the Giessen group's model is based on the solutions of the coupled-channel equations for the $MM$-tetraquark and $D\bar D$-tetraquark amplitudes $\phi_M$ and $\phi_D$, respectively, as illustrated in \fig{GBSE}. In this case there are no contributions of type $q\bar q(T_{q\q})$,  with $D\bar D$ scattering taking place only via intermediate $MM$ states. One of the features of the Giessen group's model is that it is based on a rigorous field-theoretic derivation for the $2q2\bar q$ system where all approximations can be clearly specified. Thus, following the derivation presented in  \cite{Heupel:2012ua}, the model is covariant, retains only pair-wise interactions between the quarks, and thus leads to the use of the t matrix $T_{aa'}$ corresponding to the scattering of the 4 quarks where all interactions are switched off except those within the pairs labelled by $a$ and $a'$. It can be shown (see Eq.\ (10) of \cite{Heupel:2012ua}) that
\be
T_{aa'} = T_{a} + T_{a'} + T_{a} T_{a'}  \eqn{10}
\ee
where $T_{a}$ and $T_{a'}$ are the separate two-body t matrices for the scattering of the quarks within pairs $a$ and $a'$, respectively. The first two terms on the right hand side (RHS) of \eq{10} were neglected in the derivation of \cite{Heupel:2012ua}, yet are responsible for contributions like that of the $q\bar q(T_{q\q})$ intermediate state in the Moscow group's model. Implementing a further approximation where $T_{a}$ and $T_{a'}$ are assumed to be dominated by meson and diquark pole contributions, leads to the equations of \fig{GBSE}. 

In order to achieve a unified description where all the contributions illustrated in \fig{MBSE} and \fig{GBSE} are taken into account, we
derive coupled equations similar to those of \fig{GBSE}, but where the first two terms on the RHS of \eq{10} are retained at least to first order at this stage. The resulting equations have the same form as those of  \fig{GBSE}, but with a kernel that contains additional diagrams illustrated in \fig{pots1}. Thus, the $q\bar q(T_{q\q})$ contribution is included, as well as corresponding  $q q (T_{\q\q})$ and $\q\q (T_{qq})$ contributions. In this way we unify the Moscow and Giessen approaches, and hope that the resulting unified tetraquark equations will lead to a more accurate description of a tetraquark, including an improved assessment of the relative roles played by its $D\bar D$ and $M M$ components.

\section{Derivation}

For simplicity, in Sec.\ II A we derive general tetraquark equations for the case of distinguishable quarks. Then, in  Sec.\ II B, corresponding equations for two identical quarks and two identical antiquarks are obtained by explicitly antisymmetrizing the distinguishable quark case. In Sec.\ II C, after the introduction of separable approximations for the two-body t matrices in the product term $T_a T_{a'}$ of \eq{10}, the resulting coupled channel $MM-D\bar D$ equations are recast so as to expose three-body-like states of the form $q\bar q (T_{q\q})$, $q q (T_{\q\q})$, and $\q\q (T_{qq})$. The final part of the derivation, in Sec.\ II D, is devoted to symmetrizing  the two-meson states in the formalism, as these may not have the required symmetry for the case of identical mesons.

\subsection{Four-body equations for distinguishable quarks}

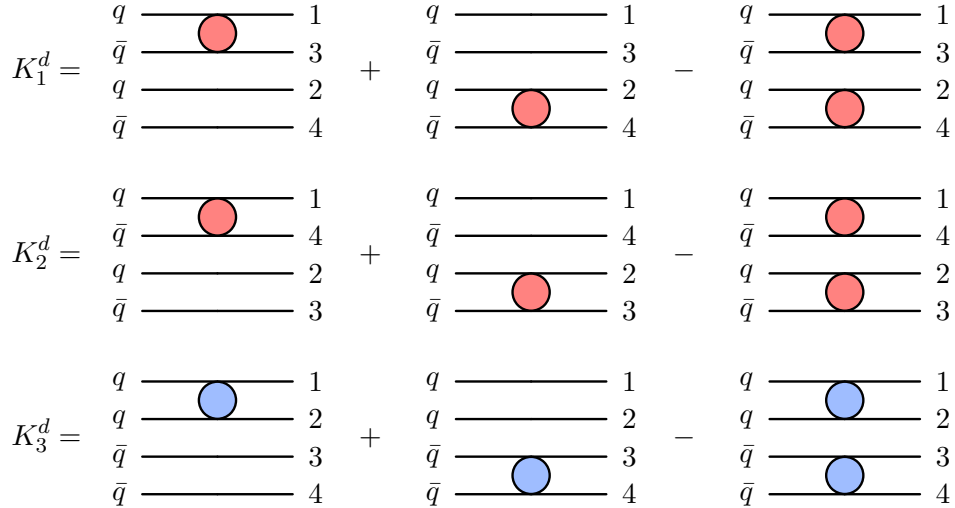
\begin{figure}[b]
\begin{center}
\begin{fmffile}{K}
\begin{align*}
\Kf_1&=\hspace{7mm}
\parbox{20mm}{
\begin{fmfgraph*}(20,15)
\fmfstraight
\fmfleft{f4,f2,f3,f1}\fmfright{i4,i2,i3,i1}
\fmf{plain,tension=1.3}{i1,v1,f1}
\fmf{plain,tension=1.3}{i2,v2,f2}
\fmf{plain,tension=1.3}{i3,v3,f3}
\fmf{plain,tension=1.3}{i4,v4,f4}
\fmffreeze
\fmf{phantom,tension=1.3}{v1,v13,v3}
\fmfv{d.s=circle,d.f=empty,d.si=14,background=(1,,.51,,.5)}{v13}
\fmf{phantom,tension=1.3}{v2,v24,v4}
\fmfv{label=$1$,l.a=0}{i1}
\fmfv{label=$2$,l.a=0}{i2}
\fmfv{label=$3$,l.a=0}{i3}
\fmfv{label=$4$,l.a=0}{i4}
\fmfv{label=$q$,l.a=180}{f1}
\fmfv{label=$q$,l.a=180}{f2}
\fmfv{label=$\q$,l.a=180}{f3}
\fmfv{label=$\q$,l.a=180}{f4}
\end{fmfgraph*}}
\hspace{8mm} +
\hspace{9mm}
\parbox{20mm}{
\begin{fmfgraph*}(20,15)
\fmfstraight
\fmfleft{f3,f2,f4,f1}\fmfright{i3,i2,i4,i1}
\fmf{plain,tension=1.3}{i1,v1,f1}
\fmf{plain,tension=1.3}{i2,v2,f2}
\fmf{plain,tension=1.3}{i3,v3,f3}
\fmf{plain,tension=1.3}{i4,v4,f4}
\fmffreeze
\fmf{phantom,tension=1.3}{v1,v14,v4}
\fmf{phantom,tension=1.3}{v2,v23,v3}
\fmfv{d.s=circle,d.f=empty,d.si=14,background=(1,,.51,,.5)}{v23}
\fmfv{label=$1$,l.a=0}{i1}
\fmfv{label=$2$,l.a=0}{i2}
\fmfv{label=$4$,l.a=0}{i3}
\fmfv{label=$3$,l.a=0}{i4}
\fmfv{label=$q$,l.a=180}{f1}
\fmfv{label=$q$,l.a=180}{f2}
\fmfv{label=$\q$,l.a=180}{f3}
\fmfv{label=$\q$,l.a=180}{f4}
\end{fmfgraph*}}
\hspace{8mm}
-
\hspace{9mm}
\parbox{20mm}{
\begin{fmfgraph*}(20,15)
\fmfstraight
\fmfleft{f4,f3,f2,f1}\fmfright{i4,i3,i2,i1}
\fmf{plain,tension=1.3}{i1,v1,f1}
\fmf{plain,tension=1.3}{i2,v2,f2}
\fmf{plain,tension=1.3}{i3,v3,f3}
\fmf{plain,tension=1.3}{i4,v4,f4}
\fmffreeze
\fmf{phantom,tension=1.3}{v1,v12,v2}
\fmfv{d.s=circle,d.f=empty,d.si=14,background=(1,,.51,,.5)}{v12}
\fmf{phantom,tension=1.3}{v3,v34,v4}
\fmfv{d.s=circle,d.f=empty,d.si=14,background=(1,,.51,,.5)}{v34}
\fmfv{label=$1$,l.a=0}{i1}
\fmfv{label=$3$,l.a=0}{i2}
\fmfv{label=$2$,l.a=0}{i3}
\fmfv{label=$4$,l.a=0}{i4}
\fmfv{label=$q$,l.a=180}{f1}
\fmfv{label=$\q$,l.a=180}{f2}
\fmfv{label=$q$,l.a=180}{f3}
\fmfv{label=$\q$,l.a=180}{f4}
\end{fmfgraph*}}    \\[8mm]
\Kf_2&=\hspace{7mm}
\parbox{20mm}{
\begin{fmfgraph*}(20,15)
\fmfstraight
\fmfleft{f4,f2,f3,f1}\fmfright{i4,i2,i3,i1}
\fmf{plain,tension=1.3}{i1,v1,f1}
\fmf{plain,tension=1.3}{i2,v2,f2}
\fmf{plain,tension=1.3}{i3,v3,f3}
\fmf{plain,tension=1.3}{i4,v4,f4}
\fmffreeze
\fmf{phantom,tension=1.3}{v1,v13,v3}
\fmfv{d.s=circle,d.f=empty,d.si=14,background=(1,,.51,,.5)}{v13}
\fmf{phantom,tension=1.3}{v2,v24,v4}
\fmfv{label=$1$,l.a=0}{i1}
\fmfv{label=$2$,l.a=0}{i2}
\fmfv{label=$4$,l.a=0}{i3}
\fmfv{label=$3$,l.a=0}{i4}
\fmfv{label=$q$,l.a=180}{f1}
\fmfv{label=$q$,l.a=180}{f2}
\fmfv{label=$\q$,l.a=180}{f3}
\fmfv{label=$\q$,l.a=180}{f4}
\end{fmfgraph*}}
\hspace{8mm} +
\hspace{9mm}
\parbox{20mm}{
\begin{fmfgraph*}(20,15)
\fmfstraight
\fmfleft{f3,f2,f4,f1}\fmfright{i3,i2,i4,i1}
\fmf{plain,tension=1.3}{i1,v1,f1}
\fmf{plain,tension=1.3}{i2,v2,f2}
\fmf{plain,tension=1.3}{i3,v3,f3}
\fmf{plain,tension=1.3}{i4,v4,f4}
\fmffreeze
\fmf{phantom,tension=1.3}{v1,v14,v4}
\fmf{phantom,tension=1.3}{v2,v23,v3}
\fmfv{d.s=circle,d.f=empty,d.si=14,background=(1,,.51,,.5)}{v23}
\fmfv{label=$1$,l.a=0}{i1}
\fmfv{label=$2$,l.a=0}{i2}
\fmfv{label=$3$,l.a=0}{i3}
\fmfv{label=$4$,l.a=0}{i4}
\fmfv{label=$q$,l.a=180}{f1}
\fmfv{label=$q$,l.a=180}{f2}
\fmfv{label=$\q$,l.a=180}{f3}
\fmfv{label=$\q$,l.a=180}{f4}
\end{fmfgraph*}}
\hspace{8mm}
-
\hspace{9mm}
\parbox{20mm}{
\begin{fmfgraph*}(20,15)
\fmfstraight
\fmfleft{f4,f3,f2,f1}\fmfright{i4,i3,i2,i1}
\fmf{plain,tension=1.3}{i1,v1,f1}
\fmf{plain,tension=1.3}{i2,v2,f2}
\fmf{plain,tension=1.3}{i3,v3,f3}
\fmf{plain,tension=1.3}{i4,v4,f4}
\fmffreeze
\fmf{phantom,tension=1.3}{v1,v12,v2}
\fmfv{d.s=circle,d.f=empty,d.si=14,background=(1,,.51,,.5)}{v12}
\fmf{phantom,tension=1.3}{v3,v34,v4}
\fmfv{d.s=circle,d.f=empty,d.si=14,background=(1,,.51,,.5)}{v34}
\fmfv{label=$1$,l.a=0}{i1}
\fmfv{label=$4$,l.a=0}{i2}
\fmfv{label=$2$,l.a=0}{i3}
\fmfv{label=$3$,l.a=0}{i4}
\fmfv{label=$q$,l.a=180}{f1}
\fmfv{label=$\q$,l.a=180}{f2}
\fmfv{label=$q$,l.a=180}{f3}
\fmfv{label=$\q$,l.a=180}{f4}
\end{fmfgraph*}}    \\[8mm]
\Kf_3&=\hspace{7mm}
\parbox{20mm}{
\begin{fmfgraph*}(20,15)
\fmfstraight
\fmfleft{f4,f2,f3,f1}\fmfright{i4,i2,i3,i1}
\fmf{plain,tension=1.3}{i1,v1,f1}
\fmf{plain,tension=1.3}{i2,v2,f2}
\fmf{plain,tension=1.3}{i3,v3,f3}
\fmf{plain,tension=1.3}{i4,v4,f4}
\fmffreeze
\fmf{phantom,tension=1.3}{v1,v13,v3}
\fmfv{d.s=circle,d.f=empty,d.si=14,background=(.6235,,.7412,,1)}{v13}
\fmf{phantom,tension=1.3}{v2,v24,v4}
\fmfv{label=$1$,l.a=0}{i1}
\fmfv{label=$3$,l.a=0}{i2}
\fmfv{label=$2$,l.a=0}{i3}
\fmfv{label=$4$,l.a=0}{i4}
\fmfv{label=$q$,l.a=180}{f1}
\fmfv{label=$\q$,l.a=180}{f2}
\fmfv{label=$q$,l.a=180}{f3}
\fmfv{label=$\q$,l.a=180}{f4}
\end{fmfgraph*}}
\hspace{8mm} +
\hspace{9mm}
\parbox{20mm}{
\begin{fmfgraph*}(20,15)
\fmfstraight
\fmfleft{f3,f2,f4,f1}\fmfright{i3,i2,i4,i1}
\fmf{plain,tension=1.3}{i1,v1,f1}
\fmf{plain,tension=1.3}{i2,v2,f2}
\fmf{plain,tension=1.3}{i3,v3,f3}
\fmf{plain,tension=1.3}{i4,v4,f4}
\fmffreeze
\fmf{phantom,tension=1.3}{v1,v14,v4}
\fmf{phantom,tension=1.3}{v2,v23,v3}
\fmfv{d.s=circle,d.f=empty,d.si=14,background=(.6235,,.7412,,1)}{v23}
\fmfv{label=$1$,l.a=0}{i1}
\fmfv{label=$3$,l.a=0}{i2}
\fmfv{label=$4$,l.a=0}{i3}
\fmfv{label=$2$,l.a=0}{i4}
\fmfv{label=$q$,l.a=180}{f1}
\fmfv{label=$\q$,l.a=180}{f2}
\fmfv{label=$\q$,l.a=180}{f3}
\fmfv{label=$q$,l.a=180}{f4}
\end{fmfgraph*}}
\hspace{8mm}
-
\hspace{9mm}
\parbox{20mm}{
\begin{fmfgraph*}(20,15)
\fmfstraight
\fmfleft{f4,f3,f2,f1}\fmfright{i4,i3,i2,i1}
\fmf{plain,tension=1.3}{i1,v1,f1}
\fmf{plain,tension=1.3}{i2,v2,f2}
\fmf{plain,tension=1.3}{i3,v3,f3}
\fmf{plain,tension=1.3}{i4,v4,f4}
\fmffreeze
\fmf{phantom,tension=1.3}{v1,v12,v2}
\fmfv{d.s=circle,d.f=empty,d.si=14,background=(.6235,,.7412,,1)}{v12}
\fmf{phantom,tension=1.3}{v3,v34,v4}
\fmfv{d.s=circle,d.f=empty,d.si=14,background=(.6235,,.7412,,1)}{v34}
\fmfv{label=$1$,l.a=0}{i1}
\fmfv{label=$2$,l.a=0}{i2}
\fmfv{label=$3$,l.a=0}{i3}
\fmfv{label=$4$,l.a=0}{i4}
\fmfv{label=$q$,l.a=180}{f1}
\fmfv{label=$q$,l.a=180}{f2}
\fmfv{label=$\q$,l.a=180}{f3}
\fmfv{label=$\q$,l.a=180}{f4}
\end{fmfgraph*}}
\end{align*}
\end{fmffile}   
\vspace{-3mm}

\caption{\fign{K}  Structure of the terms $\Kf_\A$ ($\A=1, 2, 3$) making up the four-body kernel $K^d$ where only two-body forces are included. The coloured circles represent two-body kernels $K^d_{ij}$ for the scattering of quarks $i$ and $j$, as indicated.}
\end{center}
\end{figure}
To describe the $2q2\bar q$ system where coupling to $q\bar q$ channels is neglected and only pairwise interactions are taken into account, we follow the formulation of Khvedelidze and Kvinikhidze \cite{Khvedelidze:1991qb} in the same way as in Ref.\ \cite{Heupel:2012ua} and in our previous work \cite{Kvinikhidze:2014yqa}. Thus,
assigning labels 1,2 to the quarks and 3,4 to the antiquarks, the $q\bar q$-irreducible 4-body kernel for distinguishable particles, $K^d$, is written as a sum of three terms whose structure is illustrated in \fig{K}, and correspondingly expressed as
\be
K^d=\sum_{aa'} K^{d}_{aa'} =\sum_{\A} K^{d}_{\A}  \eqn{pair}
\ee
where the index $a \in \left\{12,13,14,23,24,34\right\}$ enumerates six possible pairs of particles, 
the double index $aa'  \in \left\{(13,24), (14,23),(12,34)\right\}$ enumerates three possible two pairs of particles, and the Greek 
index $\A$ is used as an abbreviation for $aa'$ such that $\A=1$ denotes  $aa'=(13,24)$,  $\A=2$ denotes $aa'=(14,23)$, and  $\A=3$ denotes $aa'=(12,34)$.   Thus $K^{d}_\A\equiv K^{d}_{aa'}$ describes the part of
the four-body kernel where all interactions are switched off except those within the pairs $a$ and $a'$. Figure \ref{K} illustrates the fact that
 $K^{d}_\A$ can be expressed in terms of the two-body kernels $K^{d}_a$ and $K^{d}_{a'}$ as \cite{Khvedelidze:1991qb,Heupel:2012ua,Kvinikhidze:2014yqa},
\be
K^{d}_\A=K^{d}_a G_{a'}^{0}{}^{-1}+K^{d}_{a'}G_{a}^{0}{}^{-1} -K^{d}_aK^{d}_{a'}, \eqn{Kaa}
\ee
where $G_{a}^{0}$ ( $G_{a'}^{0}$ ) is the 2-body disconnected Green function for particle pair $a$ ($a'$). Of note is the presence of a minus sign in the last term of \eq{Kaa}, which is necessary to avoid overcounting.

 To simplify the notation, we shall suppress writing disconnected Green functions whenever these are self-evident; thus we may write \eq{Kaa} as the three expressions
\begin{subequations} \eqn{K1=K13}
\begin{align}
K^{d}_1& =K^{d}_{13}+K^{d}_{24}-K^{d}_{13}K^{d}_{24},\\
K^{d}_2& =K^{d}_{14}+K^{d}_{23}-K^{d}_{14}K^{d}_{23} ,\\
K^{d}_3&=K^{d}_{12}+K^{d}_{34}-K^{d}_{12}K^{d}_{34} ,
\end{align}
\end{subequations}
and the $2q2\q$ kernel for distinguishable quarks in the pairwise approximation, as
\be
K^d=K^{d}_1+K^{d}_2+K^{d}_3.
\ee
Although the superscript ``$d$'' (to indicate the distinguishable particle assumption) is redundant for quantities like $K_1^{d}$ and $K_2^{d}$ involving $q\q$ pairs, we keep it for the moment in order to avoid a mixed notation.
 
The $2q2\q$  bound state form factor for distinguishable quarks is then
\be
 \Phi^d= K^d G_0^{(4)} \Phi^d       \eqn{dist2q2q}
 \ee
where $G_0^{(4)}$ is the fully disconnected part of the full $2q2\q$ Green function $G^{(4)}$ \cite{Kvinikhidze:2021kzu}.
The four-body kernels $K_\A$ can be used to define the Faddeev components of $\Phi^d$ as
\be
\Phi^d_\A = K^{d}_\A G_0^{(4)} \Phi^d,
\ee
so that
\be
\sum_\A \Phi^d_\A = \Phi^d.
\ee
From \eq{dist2q2q} follow Faddeev-like equations for the components,
\be
\Phi^d_\alpha=T^{d}_\alpha\sum_\beta\bar\delta_{\alpha\beta}G_0^{(4)}\Phi^d_\beta    \eqn{PhidFad}
\ee
where $\bar\delta_{\alpha\beta}=1-\delta_{\alpha\beta}$ and $T^{d}_\A$ is the t matrix corresponding to kernel $\Kf_\A$; that is,
\be
 T^{d}_\A = K^{d}_\A + K^{d}_\A \Gf_0 T^{d}_\A
\ee
with $T^{d}_\A$ being expressed in terms of  two-body t matrices $T^{d}_a$ and $T^{d}_{a'}$ as
\be
T^{d}_\A=T^{d}_a G_{a'}^{0}{}^{-1}+T^{d}_{a'}G_{a}^{0}{}^{-1} + T^{d}_aT^{d}_{a'}, \eqn{Tad}
\ee
or in the simplified notation analogous to \eq{K1=K13},
\begin{subequations}\eqn{T1=T13}
\begin{align}
T^{d}_1& =T^{d}_{13}+T^{d}_{24}+T^{d}_{13}T^{d}_{24},\eqn{T1=T13a}\\
T^{d}_2& =T^{d}_{14}+T^{d}_{23}+T^{d}_{14}T^{d}_{23} , \eqn{T1=T13b}\\
T^{d}_3& =T^{d}_{12}+T^{d}_{34}+T^{d}_{12}T^{d}_{34} . \eqn{T1=T13c}
\end{align}
\end{subequations}
Equations (\ref{PhidFad}) can likewise be written {with dropped $G_0^{(4)}$'s} as
\begin{subequations} \eqn{disting-qq}
\begin{align}
\Phi^d_1&=T^{d}_1 (\Phi^d_2+\Phi^d_3), \\
\Phi^d_2&=T^{d}_2 (\Phi^d_3+\Phi^d_1), \\
\Phi^d_3&=T^{d}_3 (\Phi^d_1+\Phi^d_2) . 
\end{align}
\end{subequations}

\subsection{Four-body equations for indistinguishable quarks}

The $2q2\q$  bound state form factor $\Phi$ for two identical quarks $1,2$ and two identical antiquarks $3,4$, satisfies the equation
\be
\Phi=\frac{1}{4}KG_0^{(4)}\Phi   \eqn{ident2q2q}
\ee
where the kernel $K$ is antisymmetric with respect to swapping quark or antiquark quantum numbers either in the initial or in the final state; that is,
 \be
 {\cal P}_{34}K={\cal P}_{12}K=K{\cal P}_{34}=K{\cal P}_{12}=-K   \eqn{Kas}
 \ee
 where the exchange operator $\PP_{ij}$ swaps the quantum numbers associated with particles $i$ and $j$ in the quantity on which it is operating; for example,  ${\cal P}_{12}\Phi(p_1p_2p_3p_4)=\Phi(p_2p_1p_3p_4)$ and ${\cal P}_{34}\Phi(p_1p_2p_3p_4)=\Phi(p_1p_2p_4p_3)$.
The factor $ \frac{1}{4}$ in   \eq{ident2q2q} is a product of the combinatorial factors $\frac{1}{2}$, one for identical quarks and another for  identical antiquarks.
 The in this way antisymmetric kernel $K$ can be represented as
 \be
 K=(1-{\cal P}_{12})(1-{\cal P}_{34})K^d
 \ee
 where $K^d$ is symmetric with respect to swapping either quark or antiquark quantum numbers in the initial and final states simultaneously, ${\cal P}_{12}K^d{\cal P}_{12}={\cal P}_{34}K^d{\cal P}_{34}=K^d$. This symmetry property of $K^d$ can be written in the form of commutation relations
 \be
 [{\cal P}_{34},K^d]=[{\cal P}_{12},K^d]=0,    \eqn{Kdcom}
 \ee
and follows directly from  the following relations implied by \eqs{K1=K13}:
\begin{subequations}\eqn{Ksym}
\begin{align}  
{\cal P}_{12}K^{d}_3{\cal P}_{12}& ={\cal P}_{34}K^{d}_3{\cal P}_{34}=K^{d}_3, \\
{\cal P}_{12}K^{d}_1{\cal P}_{12}& ={\cal P}_{34}K^{d}_1{\cal P}_{34}=K^{d}_2.
\end{align}
\end{subequations}
Due to the antisymmetry properties of $K$ as specified in \eq{Kas}, the solution of the identical particle bound state equation, \eq{ident2q2q}, is correspondingly antisymmetric; namely,  ${\cal P}_{34}\Phi={\cal P}_{12}\Phi=-\Phi$.
However, because $K^d$ usually corresponds to a fewer number of diagrams than $K$, rather than solving \eq{ident2q2q}, it may be more convenient to determine $\Phi$ by antisymmetrising the solution $\Phi^d$ of the bound state equation for distinguishable quarks, as
\be
\Phi=(1-{\cal P}_{12})(1-{\cal P}_{34})\Phi^d. \eqn{PhiAS}
\ee
Then, in view of the commutation relations of \eq{Kdcom}, if the solution $\Phi^d$
 exists, its antisymmetrized version as given by \eq{PhiAS}, also satisfies the bound state equation for distinguishable quarks, \eq{dist2q2q}, as well as the one for indistinguishable ones,  \eq{ident2q2q}:
\begin{align}
{\Phi}&=(1-{\cal P}_{12})(1-{\cal P}_{34})K^dG_0^{(4)}\Phi ^d \nn
& =K^dG_0^{(4)}(1-{\cal P}_{12})(1-{\cal P}_{34})\Phi ^d =K^dG_0^{(4)}\Phi  \nn
&= \frac{1}{4}K^dG_0^{(4)}(1-{\cal P}_{12})(1-{\cal P}_{34})\Phi \nn
&=\frac{1}{4}K G_0^{(4)}\Phi.  \eqn{11**}
\end{align}
A further consequence of the commutation relations of \eq{Kdcom}, is that the system corresponding to the kernel $K^d$ is degenerate, having multiple linearly independent solutions (eigenfunctions) corresponding to one eigenenergy (tetraquark mass), unless by chance $K^d$ is symmetric or antisymmetric in the final and initial state variables independently,  ${\cal P}_{34}K^d={\cal P}_{12}K^d=\pm K^d$.
In the case of the $2q 2\bar q$ system, there are 4 such eigenfunctions related to each other by quark swapping operators, or symmetrized in 4 possible ways using $(1\pm{\cal P}_{12})$ and $(1\pm{\cal P}_{34})$. By contrast, the system corresponding to the kernel $K$ is {\it not} degenerate, because $K$ is fully antisymmetric from both (initial and final state) sides independently and consequently
it has only one, fully antisymmetric, solution. Indeed, in this system any swapping of the identical-quark quantum numbers does not change the wave function because only fully antisymmetric wave functions satisfy the bound state equation, ${\cal P}_{ij}\Phi=\frac{1}{4}{\cal P}_{ij}K G_0^{(4)}\Phi=-\frac{1}{4}K G_0^{(4)}\Phi=-\Phi$.

As $\Phi$ satisfies the same bound state equation as $\Phi^d$,
\be
\Phi= K^d G_0^{(4)} \Phi ,  \eqn{PhiBS}
\ee
the kernels $K^{d}_\A$ can again be used to define Faddeev components, but this time for $\Phi$:
\be
\Phi_\A = K^{d}_\A G_0^{(4)} \Phi,
\ee
where
\be
\sum_\A \Phi_\A = \Phi.
\ee
In view of \eqs{Ksym}, the Faddeev components $\Phi_\alpha$  have the following properties:
\begin{subequations} \eqn{P12Phi1= P34Phi1}\eqn{Phisym}
\begin{alignat}{2}
{\cal P}_{12}\Phi_3 & =-\Phi_3, & \hspace{1cm}
{\cal P}_{12}\Phi_1 & =-\Phi_2, \eqn{Phisyma}\\
{\cal P}_{34}\Phi_3 & =-\Phi_3, & \hspace{1cm}
{\cal P}_{34}\Phi_1 &=-\Phi_2.   \eqn{Phisymb}
\end{alignat}
\end{subequations}

Since $\Phi$ satisfies the same bound state equation as $\Phi^d$,  the components $\Phi_\A$ satisfy the same Faddeev-like equations as for distinguishable quarks [\eqs{disting-qq}], 
\begin{subequations}  \eqn{distj}
\begin{align}
\Phi_1&=T^{d}_1 (\Phi_2+\Phi_3), \eqn{distij}\\
\Phi_2&=T^{d}_2 (\Phi_3+\Phi_1) ,\eqn{distjb}\\
\Phi_3&=T^{d}_3 (\Phi_1+\Phi_2)   . \eqn{distjc}
\end{align}
\end{subequations}
{where, like in \eqs{disting-qq}, factors of $G_0^{(4)}$ have been dropped.}
Although an arbitrary solution  $\{\Phi_1, \Phi_2,\Phi_3\}$ of  \eqs{disti} will not necessarily obey the symmetry properties of \eq{P12Phi1= P34Phi1}, we note that if $\{\Phi_1, \Phi_2,\Phi_3\}$ is a solution, then so is $\PP_{12}\{\Phi_2, \Phi_1,\Phi_3\}$ and  $\PP_{34}\{\Phi_2, \Phi_1,\Phi_3\}$, and therefore so are their linear combinations
\begin{align}
\{\Phi'_1, \Phi'_2,\Phi'_3\} &=\{\Phi_1, \Phi_2,\Phi_3\} -\PP_{12}\{\Phi_2, \Phi_1,\Phi_3\},
\end{align}
and
\be
\{\Phi''_1, \Phi''_2,\Phi''_3\} =\{\Phi'_1, \Phi'_2,\Phi'_3\} -\PP_{34}\{\Phi'_2, \Phi'_1,\Phi'_3\}
\ee
which does have these symmetry properties. Thus, without loss of generality,  we shall assume that we are dealing with a solution $\{\Phi_1, \Phi_2,\Phi_3\}$  of \eqs{distj} which has the symmetry properties of \eq{P12Phi1= P34Phi1}. We also note that the input 2-body t matrices $T^{d}_{12}$ and $T^{d}_{34}$ can be antisymmetrized by defining
\be
T_{12} = \frac{1}{2}(1-\PP_{12}) T^{d}_{12}, \hspace{5mm}
T_{34} = \frac{1}{2}(1-\PP_{34}) T^{d}_{34},
\ee
so that
\begin{subequations}  \eqn{Tanti}
\begin{align}
T_{12} \PP_{12} &= \PP_{12} T_{12} = - T_{12} ,\\
T_{34} \PP_{34} &= \PP_{34} T_{34} = - T_{34},
\end{align}
\end{subequations}
which also allows \eq{Tad} to be extended to the case of identical particles as
\be
T_\A=T_a G_{a'}^{0}{}^{-1}+T_{a'}G_{a}^{0}{}^{-1} + T_aT_{a'}, \eqn{Ta}
\ee
or explicitly with $G_{a'}^{0}{}^{-1}$ and $G_{a}^{0}{}^{-1}$ suppressed,
\begin{subequations}\eqn{T3=T12}
\begin{align}
T_1& =T_{13}+T_{24}+T_{13}T_{24},\eqn{T3=T12a}\\
T_2& =T_{14}+T_{23}+T_{14}T_{23} , \eqn{T3=T12b}\\
T_3& =T_{12}+T_{34}+T_{12}T_{34} , \eqn{T3=T12c}
\end{align}
\end{subequations}
where the equations for $T_1$ and $T_2$ are just those of \eq{T1=T13a} and  \eq{T1=T13b} written without the redundant ``$d$"  superscripts, and where $T_3$ is defined by \eq{T3=T12c}. Furthermore, as the physical (antisymmetric) t matrices for $qq$ and $\q\q$ scattering are $T_{qq} = (1-{\cal P}_{12})T^d_{12} =2T_{12}$ and $T_{\q\q} = (1-{\cal P}_{34})T^d_{34} =2T_{34}$, respectively, it is convenient to use the antisymmetric $T_{12}$ and $T_{34}$ as the input $qq$ and $\q\q$ t matrices. This is accomplished by multiplying \eq{distjc} by $(1-\PP_{12})$ and using the symmetry properties of \eq{P12Phi1= P34Phi1} to obtain
\begin{align}
\Phi_3 &=\frac{1}{2}(1-\PP_{12}) T^{d}_3 \frac{1}{2}(1-\PP_{34}) (\Phi_1+\Phi_2)\nn
&=T_3 (\Phi_1+\Phi_2)
\end{align}
thereby allowing us to write \eqs{distj} as
\begin{subequations}  \eqn{disti}
\begin{align}
\Phi_1&=T_1 (\Phi_2+\Phi_3), \eqn{distia}\\
\Phi_2&=T_2 (\Phi_3+\Phi_1) ,\eqn{distib}\\
\Phi_3&=T_3 (\Phi_1+\Phi_2)   . \eqn{distic}
\end{align}
\end{subequations}

For physical (antisymmetric) solutions of  \eqs{disti}, only two of these three equations are independent. For example,  \eq{distib} can be written as
\begin{align}
-{\cal P}_{12}\Phi_1&={\cal P}_{12}T_1{\cal P}_{12} (\Phi_3+\Phi_1) \nn
&={\cal P}_{12}T_1 (-\Phi_3 - \Phi_2)
\end{align}
where \eq{P12Phi1= P34Phi1} and $T_2={\cal P}_{12}T_1{\cal P}_{12}$ have been used; then, after a further application of ${\cal P}_{12}$, one obtains  \eq{distia}. Choosing \eq{distia} and \eq{distic} as the two independent equations, we can use $ \Phi_2 = -{\cal P}_{12}\Phi_1$ to obtain closed equations
\begin{subequations}  \eqn{1,3}
\begin{align}
\Phi_1&=T_1 (-{\cal P}_{12}\Phi_1+\Phi_3), \eqn{1,3a}\\
\Phi_3&=T_3 (\Phi_1-{\cal P}_{12}\Phi_1), \eqn{1,3b}
\end{align}
\end{subequations}
where, necessarily, ${\cal P}_{12}\Phi_3  = -\Phi_3$. In this way an arbitrary solution of \eqs{1,3} results in components $\{\Phi_1, \Phi_2,\Phi_3\}$ which obey the symmetry properties of \eqs{Phisyma} but not necessarily of \eq{Phisymb}; however, invoking a similar argument as previously, no generality is lost in choosing a solution of \eqs{1,3} that has all the symmetry properties of \eq{P12Phi1= P34Phi1}.

Equation (\ref{1,3b}) can be further simplified using ${\cal P}_{12}\Phi_1={\cal P}_{34}\Phi_1$  and the assumption that $T_{12}$ and $T_{34}$ are antisymmetric in their labels,  so that
\begin{align}
T_3 {\cal P}_{12}\Phi_1 &= (T_{12}+T_{34}+T_{12}T_{34}){\cal P}_{12}\Phi_1 =-T_3 \Phi_1.
\end{align}
In this way \eqs{1,3} take the form
\begin{subequations}\eqn{1,3****}
\begin{align}
\Phi_1&=T_1 (-{\cal P}_{12}\Phi_1+\Phi_3) \eqn{1,3a****} \\
\Phi_3&=2T_3 \Phi_1. \eqn{1,3b****}
\end{align}
\end{subequations}
Again, without loss of generality,  we choose a solution of \eqs{1,3****} which has all the symmetry properties of \eq{P12Phi1= P34Phi1}.

\subsection{Tetraquark equations with exposed  $\bm{q\bar q (T_{q\q})}$, $\bm{q q (T_{\q\q})}$, and $\bm{\q\q (T_{qq})}$ channels}\label{dd-int}

Choosing \eqs{1,3****} as the four-body equations describing a tetraquark, they may be expressed in matrix form as
\be
\PPhi = \TT\RR\PPhi       \eqn{P,12}
\ee
where
{
\be
\PPhi = \begin{pmatrix} \Phi_1 \\  \Phi_3 \end{pmatrix} ,\hspace{2mm}
\TT = \begin{pmatrix} \frac{1}{2} T_1  & 0\\  0&T_3 \end{pmatrix}  ,\hspace{2mm}
\RR =    2\begin{pmatrix} -  {\cal P}_{12} & 1 \\  1 &0 \end{pmatrix} .
\ee}
Writing
\be
T_1 =T^\times_1+T^+_1 ,\hspace{1cm}
T_3 =T^\times_3+T^+_3 ,
\ee
where
\begin{subequations} \eqn{Tx}
\begin{alignat}{2}
T^\times_1&=T_{13}T_{24}, & \hspace{1cm} T^+_1&=T_{13}+T_{24}, \eqn{Txa}\\
T^\times_3& =T_{12}T_{34} & \hspace{1cm} T^+_3&=T_{12}+T_{34}, \eqn{Txb}
\end{alignat}
\end{subequations}
we have that
\be
\TT = \TT^\times + \TT^+
\ee
where
{
\be
\TT^\times= \begin{pmatrix} \frac{1}{2} T^\times_1  & 0\\  0&T^\times_3 \end{pmatrix}  ,\hspace{2mm}
\TT^+ = \begin{pmatrix} \frac{1}{2}T^+_1  & 0\\  0&T^+_3 \end{pmatrix}  .
\ee}
Thus
\be
\PPhi = (\TT^\times + \TT^+) \RR\PPhi 
\ee
and consequently
\be
\PPhi = (1- \TT^+\RR)^{-1} \TT^\times \RR\PPhi  .  \eqn{(-1)}
\ee

To be close to previous publications we choose a separable approximation for the two-body t matrices in $T^\times_1$ and $T^\times_3$ (but not necessarily in $T^+_1$ and $T^+_3$); namely, for $a \in \left\{13, 24,12, 34\right\}$ we take
\be
T_{a}=i\Gamma_{a} D_{a} \bar\Gamma_{a},
\ee
where $D_{a}=D_{a}(P_{a})$ is a propagator whose structure can be chosen to best describe the two-body t matrix $T_{a}$, and $\Gamma_{a}$ is a corresponding vertex function.  In the simplest case, one can follow previous publications and choose the pole approximation  where $D_{a}(P_{a})=1/(P_{a}^2-m_{a}^2)$ is the propagator for the bound particle (diquark, antidiquark, or meson) of mass $m_a$. In view of \eq{Tanti}, note that
\begin{subequations}
\begin{alignat}{2}
\PP_{12} \Gamma_{12} &= - \Gamma_{12}, & \hspace{1cm}  \bar\Gamma_{12} \PP_{12}&= - \bar\Gamma_{12}, \\
\PP_{34} \Gamma_{34} &= - \Gamma_{34}, & \hspace{1cm}  \bar\Gamma_{34} \PP_{34}&= - \bar\Gamma_{34}.
\end{alignat}
\end{subequations}
We can thus write
\be
\TT^\times = -\GG D\bGG     \eqn{Txsep}
\ee
where
{
\be
\GG= \begin{pmatrix} \Gamma_{13}\Gamma_{24}  & 0\\  0& \Gamma_{12}\Gamma_{34} \end{pmatrix}  ,\hspace{2mm}
D = \begin{pmatrix}  \frac{1}{2}D_{13}D_{24}  & 0\\  0& D_{12} D_{34} \end{pmatrix}  ,\hspace{2mm}
\bGG= \begin{pmatrix} \bar \Gamma_{13}\bar\Gamma_{24}  & 0\\  0& \bar\Gamma_{12}\bar\Gamma_{34} \end{pmatrix} .
\ee}
In this way $\TT^\times$ exposes intermediate state meson-meson $(D_{13}D_{24})$ and diquark-antidiquark $(D_{12}D_{34})$ channels. Using \eq{Txsep} in \eq{(-1)}, 
\be
\phi =-\bGG \RR (1- \TT^+\RR)^{-1} \GG D\phi  \eqn{2-step-eq}
\ee
where
\be
\phi = \bGG \RR\PPhi .  \eqn{2-step-phi}
\ee
In this way we obtain the bound state equation for $\phi$ in meson-meson $(MM)$ and diquark-antidiquark $(D\bar D)$  space,
\be 
\phi=V D\phi    \eqn{400-}
\ee
where the $2\times 2$ matrix potential (with reinserted $\Gf_0 $) is
\be
V =-\bGG \RR\Gf_0 (1- \TT^+\RR\Gf_0)^{-1} \GG.    \eqn{Vfull}
\ee
Expanding the term in square brackets in powers of  $\TT^+$ (i.e., with respect to the contribution of intermediate states $q\bar q (T_{q\q})$, $q q (T_{\q\q})$, and $\q\q (T_{qq})$),
\be
V = -\bGG  \RR\Gf_0 \left[1+ \TT^+\RR\Gf_0+\dots\right]\GG,   \eqn{V}
\ee
it turns out that each of the first two terms of this expansion corresponds to different existing approaches to modelling tetraquarks in terms of $MM-D\bar D$ coupled channels. In particular, the lowest order term
\begin{align}
V^{(0)} &=  -\bGG \RR \Gf_0 \GG  \nn
&=- 2\begin{pmatrix} \bar\Gamma_1 & 0\\  0&\bar\Gamma_3 \end{pmatrix}
\begin{pmatrix} -  {\cal P}_{12} & 1 \\  1 &0 \end{pmatrix} \Gf_0 
  \begin{pmatrix}\Gamma_1  & 0\\  0&\Gamma_3 \end{pmatrix}\nn[2mm]
 &= - 2\begin{pmatrix} -  \bGamma_1\PP_{12} \Gamma_1& \bar\Gamma_1\Gamma_3\\  
  \bar\Gamma_3  \Gamma_1&0 \end{pmatrix}, \eqn{V0mat}
\end{align}
where
\begin{subequations}
\begin{alignat}{2}
\bGamma_1  &= \bGamma_{13}\bGamma_{24}, & \hspace{1cm}  \Gamma_1  &= \Gamma_{13}\Gamma_{24}, \\
\bGamma_3  &= \bGamma_{12}\bGamma_{34}, & \hspace{1cm}  \Gamma_3  &= \Gamma_{12}\Gamma_{34},
\end{alignat}
\end{subequations}
consists of Feynman diagrams illustrated in \fig{pots0}, and corresponds to the Giessen  group (GG) model of Heupel {\it et al.} \cite{Heupel:2012ua} where tetraquarks are modelled by solving the equation
\be
\phi ^{(0)}= V^{(0)} D\phi^{(0)}.
\ee
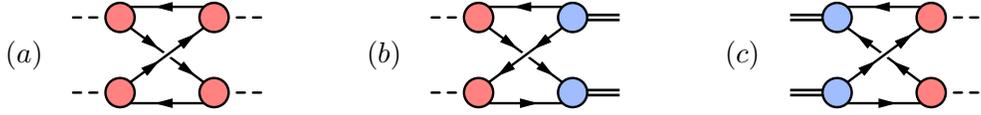
\begin{figure}[t]
\begin{center}
\begin{fmffile}{V0}
\begin{align*}
(a)\hspace{4mm}
\parbox{25mm}{
\begin{fmfgraph*}(25,15)
\fmfstraight
\fmfleftn{l}{7}\fmfrightn{r}{7}\fmfbottomn{b}{5}\fmftopn{t}{5}
\fmf{phantom}{l2,vb1,mb1,vb2,r2}
\fmf{phantom}{l6,vt1,mt1,vt2,r6}
\fmf{phantom}{l2,xb1,mb1,xb2,r2}
\fmf{phantom}{l6,xt1,mt1,xt2,r6}
\fmf{phantom}{l2,yb1,mb1,yb2,r2}
\fmf{phantom}{l6,yt1,mt1,yt2,r6}
\fmf{phantom}{t3,c,b3}
\fmffreeze
\fmfshift{4 right}{xb1}
\fmfshift{4 down}{xb1}
\fmfshift{4 left}{xt2}
\fmfshift{4 down}{xt2}
\fmfshift{4 right}{xt1}
\fmfshift{4 down}{xt1}
\fmfshift{4 left}{xb2}
\fmfshift{4 down}{xb2}
\fmfshift{4 right}{yb1}
\fmfshift{4 up}{yb1}
\fmfshift{4 left}{yt2}
\fmfshift{4 up}{yt2}
\fmfshift{4 right}{yt1}
\fmfshift{4 up}{yt1}
\fmfshift{4 left}{yb2}
\fmfshift{4 up}{yb2}
\fmfv{d.sh=circle,d.f=empty,d.si=11,background=(1,,.51,,.5)}{vb1}
\fmfv{d.sh=circle,d.f=empty,d.si=11,background=(1,,.51,,.5)}{vb2}
\fmfv{d.sh=circle,d.f=empty,d.si=11,background=(1,,.51,,.5)}{vt1}
\fmfv{d.sh=circle,d.f=empty,d.si=11,background=(1,,.51,,.5)}{vt2}
\fmf{dashes}{l2,vb1}
\fmf{dashes}{l6,vt1}
\fmf{dashes}{r2,vb2}
\fmf{dashes}{r6,vt2}
\fmfset{arrow_len}{2.5mm}
\fmf{plain,rubout=3}{yb1,xt2}
\fmf{plain}{xt1,yb2}
\fmf{phantom_arrow}{xt1,c}
\fmf{phantom_arrow}{c,yb2}
\fmf{phantom_arrow,rubout=2}{yb1,c}
\fmf{phantom_arrow,rubout=2}{c,xt2}
\fmf{fermion}{yt2,yt1}
\fmf{fermion}{xb2,xb1}
\end{fmfgraph*}}
\hspace{14mm} (b) \hspace{4mm}
\parbox{25mm}{
\begin{fmfgraph*}(25,15)
\fmfstraight
\fmfleftn{l}{7}\fmfrightn{r}{7}\fmfbottomn{b}{5}\fmftopn{t}{5}
\fmf{phantom}{l2,vb1,mb1,vb2,r2}
\fmf{phantom}{l6,vt1,mt1,vt2,r6}
\fmf{phantom}{l2,xb1,mb1,xb2,r2}
\fmf{phantom}{l6,xt1,mt1,xt2,r6}
\fmf{phantom}{l2,yb1,mb1,yb2,r2}
\fmf{phantom}{l6,yt1,mt1,yt2,r6}
\fmf{phantom}{t3,c,b3}
\fmffreeze
\fmfshift{4 right}{xb1}
\fmfshift{4 down}{xb1}
\fmfshift{4 left}{xt2}
\fmfshift{4 down}{xt2}
\fmfshift{4 right}{xt1}
\fmfshift{4 down}{xt1}
\fmfshift{4 left}{xb2}
\fmfshift{4 down}{xb2}
\fmfshift{4 right}{yb1}
\fmfshift{4 up}{yb1}
\fmfshift{4 left}{yt2}
\fmfshift{4 up}{yt2}
\fmfshift{4 right}{yt1}
\fmfshift{4 up}{yt1}
\fmfshift{4 left}{yb2}
\fmfshift{4 up}{yb2}
\fmfv{d.sh=circle,d.f=empty,d.si=11,background=(1,,.51,,.5)}{vb1}
\fmfv{d.sh=circle,d.f=empty,d.si=11,background=(.6235,,.7412,,1)}{vb2}
\fmfv{d.sh=circle,d.f=empty,d.si=11,background=(1,,.51,,.5)}{vt1}
\fmfv{d.sh=circle,d.f=empty,d.si=11,background=(.6235,,.7412,,1)}{vt2}
\fmf{dashes}{l2,vb1}
\fmf{dashes}{l6,vt1}
\fmf{dbl_plain}{r2,vb2}
\fmf{dbl_plain}{r6,vt2}
\fmfset{arrow_len}{2.5mm}
\fmf{plain,rubout=3}{xt2,yb1}
\fmf{plain}{yb2,xt1}
\fmf{phantom_arrow,rubout=2}{xt2,c}
\fmf{phantom_arrow,rubout=2}{c,yb1}
\fmf{phantom_arrow}{xt1,c}
\fmf{phantom_arrow}{c,yb2}
\fmf{fermion}{yt2,yt1}
\fmf{fermion}{xb1,xb2}
\end{fmfgraph*}}
\hspace{14mm}(c) \hspace{4mm}
\parbox{25mm}{
\begin{fmfgraph*}(25,15)
\fmfstraight
\fmfleftn{l}{7}\fmfrightn{r}{7}\fmfbottomn{b}{5}\fmftopn{t}{5}
\fmf{phantom}{l2,vb1,mb1,vb2,r2}
\fmf{phantom}{l6,vt1,mt1,vt2,r6}
\fmf{phantom}{l2,xb1,mb1,xb2,r2}
\fmf{phantom}{l6,xt1,mt1,xt2,r6}
\fmf{phantom}{l2,yb1,mb1,yb2,r2}
\fmf{phantom}{l6,yt1,mt1,yt2,r6}
\fmf{phantom}{t3,c,b3}
\fmffreeze
\fmfshift{4 right}{xb1}
\fmfshift{4 down}{xb1}
\fmfshift{4 left}{xt2}
\fmfshift{4 down}{xt2}
\fmfshift{4 right}{xt1}
\fmfshift{4 down}{xt1}
\fmfshift{4 left}{xb2}
\fmfshift{4 down}{xb2}
\fmfshift{4 right}{yb1}
\fmfshift{4 up}{yb1}
\fmfshift{4 left}{yt2}
\fmfshift{4 up}{yt2}
\fmfshift{4 right}{yt1}
\fmfshift{4 up}{yt1}
\fmfshift{4 left}{yb2}
\fmfshift{4 up}{yb2}
\fmfv{d.sh=circle,d.f=empty,d.si=11,background=(1,,.51,,.5)}{vb2}
\fmfv{d.sh=circle,d.f=empty,d.si=11,background=(.6235,,.7412,,1)}{vb1}
\fmfv{d.sh=circle,d.f=empty,d.si=11,background=(1,,.51,,.5)}{vt2}
\fmfv{d.sh=circle,d.f=empty,d.si=11,background=(.6235,,.7412,,1)}{vt1}
\fmf{dashes}{r2,vb2}
\fmf{dashes}{r6,vt2}
\fmf{dbl_plain}{l2,vb1}
\fmf{dbl_plain}{l6,vt1}
\fmfset{arrow_len}{2.5mm}
\fmf{plain,rubout=3}{yb1,xt2}
\fmf{plain}{yb2,xt1}
\fmf{phantom_arrow,rubout=2}{yb1,c}
\fmf{phantom_arrow,rubout=2}{c,xt2}
\fmf{phantom_arrow}{yb2,c}
\fmf{phantom_arrow}{c,xt1}
\fmf{fermion}{yt2,yt1}
\fmf{fermion}{xb1,xb2}
\end{fmfgraph*}}
\end{align*}
\end{fmffile}   
\vspace{-3mm}

\caption{\fign{pots0}  Feynman diagrams making up the elements of the coupled channel $MM-D\bar D$ kernel matrix $V^{(0)}$ of \eq{V0mat}: (a) $\bGamma_1\PP_{12}\Gamma_1$, (b) $\bGamma_1\Gamma_3$, and (c) $\bGamma_3\Gamma_1$. Solid lines with leftward (rightward) arrows represent quarks (antiquarks), dashed lines represent mesons, and double-lines represent diquarks and antidiquarks.}
\end{center}
\end{figure}
Similarly,  the first order correction (without the lowest order term included) is
\begin{align}
V^{(1)} &= -\bGG \RR  \Gf_0 \TT^+\RR\Gf_0\GG \nn
&=- 4\begin{pmatrix} \bar\Gamma_1 & 0\\  0&\bar\Gamma_3 \end{pmatrix}  \Gf_0
\begin{pmatrix} -  {\cal P}_{12} & 1 \\  1 &0 \end{pmatrix}  \begin{pmatrix} \frac{1}{2} T^+_1  & 0\\  0&T^+_3 \end{pmatrix} 
   \begin{pmatrix} -  {\cal P}_{12} & 1 \\  1 &0 \end{pmatrix} \Gf_0  
  \begin{pmatrix}\Gamma_1  & 0\\  0&\Gamma_3 \end{pmatrix} 
  \nn[2mm]
  &=-  2\begin{pmatrix} \bar\Gamma_1[{\cal P}_{12}  T^+_1 {\cal P}_{12}+2T^+_3]\Gamma_1 &-\bar\Gamma_1{\cal P}_{12}  T^+_1\Gamma_3 \\ 
   -\bar\Gamma_3T^+_1{\cal P}_{12}\Gamma_1 &2\bar\Gamma_3T^+_1 \Gamma_3\end{pmatrix} ,  \eqn{V1mat}
\end{align}
which consists of Feynman diagrams illustrated in \fig{pots1}, and corresponds to the Moscow group (MG) model of Faustov {\it et al.} \cite{Faustov:2020qfm} where they modelled tetraquarks by solving the equation
\be
\phi^{(1)}  = V^{(1)} D \phi^{(1)} ,
\ee
albeit, with only diquark-antidiquark channels retained. It is an essential result of this paper, that it is the sum of the potentials $V^{(0)}$ and $V^{(1)}$, each associated with the separate approaches of the Giessen and Moscow groups, with tetraquarks modelled by the bound state equation
\be
\phi = [V^{(0)}+V^{(1)}] D \phi,  \eqn{VG+VF}
\ee
that results in a complete $MM-D\bar D$ coupled channel description up to first order in  $\TT^+$ [i.e., up to first order in intermediate states where one $2q$ pair ($qq$, $q\q$, or $\q\q$) is mutually interacting while the other $2q$ pair is spectating].
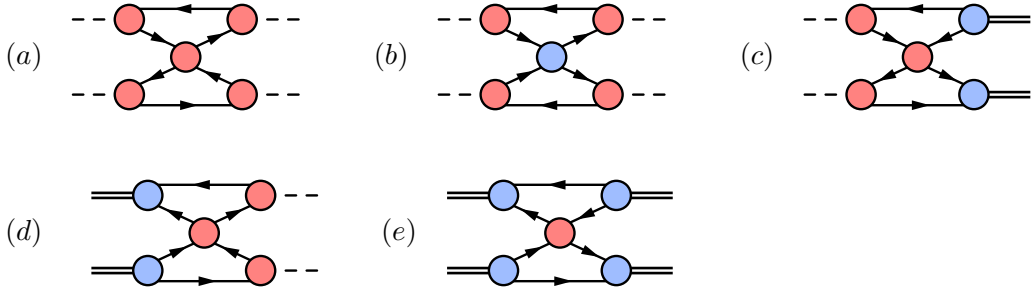
\begin{figure}[t]
\begin{center}
\begin{fmffile}{V1}
\begin{align*}
&(a)\hspace{4mm}
\parbox{30mm}{
\begin{fmfgraph*}(30,15)
\fmfstraight
\fmfleftn{l}{7}\fmfrightn{r}{7}\fmfbottomn{b}{7}\fmftopn{t}{7}
\fmf{phantom}{l2,vb1,mb1,vb2,r2}
\fmf{phantom}{l6,vt1,mt1,vt2,r6}
\fmf{phantom}{l2,xb1,mb1,xb2,r2}
\fmf{phantom}{l6,xt1,mt1,xt2,r6}
\fmf{phantom}{l2,yb1,mb1,yb2,r2}
\fmf{phantom}{l6,yt1,mt1,yt2,r6}
\fmf{phantom}{l4,c,r4}
\fmf{phantom}{l4,clb,r4}
\fmf{phantom}{l4,clt,r4}
\fmf{phantom}{l4,crb,r4}
\fmf{phantom}{l4,crt,r4}
\fmffreeze
\fmfshift{4 right}{xb1}
\fmfshift{4 down}{xb1}
\fmfshift{4 left}{xt2}
\fmfshift{4 down}{xt2}
\fmfshift{4 right}{xt1}
\fmfshift{4 down}{xt1}
\fmfshift{4 left}{xb2}
\fmfshift{4 down}{xb2}
\fmfshift{4 right}{yb1}
\fmfshift{4 up}{yb1}
\fmfshift{4 left}{yt2}
\fmfshift{4 up}{yt2}
\fmfshift{4 right}{yt1}
\fmfshift{4 up}{yt1}
\fmfshift{4 left}{yb2}
\fmfshift{4 up}{yb2}
\fmfshift{4 left}{clb}
\fmfshift{4 down}{clb}
\fmfshift{4 left}{clt}
\fmfshift{4 up}{clt}
\fmfshift{4 right}{crb}
\fmfshift{4 down}{crb}
\fmfshift{4 right}{crt}
\fmfshift{4 up}{crt}
\fmfv{d.sh=circle,d.f=empty,d.si=11,background=(1,,.51,,.5)}{vb1}
\fmfv{d.sh=circle,d.f=empty,d.si=11,background=(1,,.51,,.5)}{vb2}
\fmfv{d.sh=circle,d.f=empty,d.si=11,background=(1,,.51,,.5)}{vt1}
\fmfv{d.sh=circle,d.f=empty,d.si=11,background=(1,,.51,,.5)}{vt2}
\fmfv{d.sh=circle,d.f=empty,d.si=11,background=(1,,.51,,.5)}{c}
\fmf{dashes}{l2,vb1}
\fmf{dashes}{l6,vt1}
\fmf{dashes}{r2,vb2}
\fmf{dashes}{r6,vt2}
\fmfset{arrow_len}{2.5mm}
\fmfi{fermion}{ vloc(__crt).. vloc(__xt2)}
\fmfi{fermion}{vloc(__xt1) ..vloc(__clt)}
\fmfi{fermion}{vloc(__yb2) ..vloc(__crb)}
\fmfi{fermion}{vloc(__clb) ..vloc(__yb1)}
\fmfi{fermion}{vloc(__yt2) .. vloc(__yt1)}
\fmfi{fermion}{vloc(__xb1) .. vloc(__xb2)}
\end{fmfgraph*}}
\hspace{10mm}(b) \hspace{4mm}
\parbox{30mm}{
\begin{fmfgraph*}(30,15)
\fmfstraight
\fmfleftn{l}{7}\fmfrightn{r}{7}\fmfbottomn{b}{7}\fmftopn{t}{7}
\fmf{phantom}{l2,vb1,mb1,vb2,r2}
\fmf{phantom}{l6,vt1,mt1,vt2,r6}
\fmf{phantom}{l2,xb1,mb1,xb2,r2}
\fmf{phantom}{l6,xt1,mt1,xt2,r6}
\fmf{phantom}{l2,yb1,mb1,yb2,r2}
\fmf{phantom}{l6,yt1,mt1,yt2,r6}
\fmf{phantom}{l4,c,r4}
\fmf{phantom}{l4,clb,r4}
\fmf{phantom}{l4,clt,r4}
\fmf{phantom}{l4,crb,r4}
\fmf{phantom}{l4,crt,r4}
\fmffreeze
\fmfshift{4 right}{xb1}
\fmfshift{4 down}{xb1}
\fmfshift{4 left}{xt2}
\fmfshift{4 down}{xt2}
\fmfshift{4 right}{xt1}
\fmfshift{4 down}{xt1}
\fmfshift{4 left}{xb2}
\fmfshift{4 down}{xb2}
\fmfshift{4 right}{yb1}
\fmfshift{4 up}{yb1}
\fmfshift{4 left}{yt2}
\fmfshift{4 up}{yt2}
\fmfshift{4 right}{yt1}
\fmfshift{4 up}{yt1}
\fmfshift{4 left}{yb2}
\fmfshift{4 up}{yb2}
\fmfshift{4 left}{clb}
\fmfshift{4 down}{clb}
\fmfshift{4 left}{clt}
\fmfshift{4 up}{clt}
\fmfshift{4 right}{crb}
\fmfshift{4 down}{crb}
\fmfshift{4 right}{crt}
\fmfshift{4 up}{crt}
\fmfv{d.sh=circle,d.f=empty,d.si=11,background=(1,,.51,,.5)}{vb1}
\fmfv{d.sh=circle,d.f=empty,d.si=11,background=(1,,.51,,.5)}{vb2}
\fmfv{d.sh=circle,d.f=empty,d.si=11,background=(1,,.51,,.5)}{vt1}
\fmfv{d.sh=circle,d.f=empty,d.si=11,background=(1,,.51,,.5)}{vt2}
\fmfv{d.sh=circle,d.f=empty,d.si=11,background=(.6235,,.7412,,1)}{c}
\fmf{dashes}{l2,vb1}
\fmf{dashes}{l6,vt1}
\fmf{dashes}{r2,vb2}
\fmf{dashes}{r6,vt2}
\fmfset{arrow_len}{2.5mm}
\fmfi{fermion}{vloc(__crt) .. vloc(__xt2)}
\fmfi{fermion}{vloc(__xt1) ..vloc(__clt)}
\fmfi{fermion}{vloc(__crb) ..vloc(__yb2)}
\fmfi{fermion}{vloc(__yb1) ..vloc(__clb)}
\fmfi{fermion}{vloc(__yt2) .. vloc(__yt1)}
\fmfi{fermion}{vloc(__xb2) .. vloc(__xb1)}
\end{fmfgraph*}}
\hspace{10mm}(c) \hspace{4mm}
\parbox{30mm}{
\begin{fmfgraph*}(30,15)
\fmfstraight
\fmfleftn{l}{7}\fmfrightn{r}{7}\fmfbottomn{b}{7}\fmftopn{t}{7}
\fmf{phantom}{l2,vb1,mb1,vb2,r2}
\fmf{phantom}{l6,vt1,mt1,vt2,r6}
\fmf{phantom}{l2,xb1,mb1,xb2,r2}
\fmf{phantom}{l6,xt1,mt1,xt2,r6}
\fmf{phantom}{l2,yb1,mb1,yb2,r2}
\fmf{phantom}{l6,yt1,mt1,yt2,r6}
\fmf{phantom}{l4,c,r4}
\fmf{phantom}{l4,clb,r4}
\fmf{phantom}{l4,clt,r4}
\fmf{phantom}{l4,crb,r4}
\fmf{phantom}{l4,crt,r4}
\fmffreeze
\fmfshift{4 right}{xb1}
\fmfshift{4 down}{xb1}
\fmfshift{4 left}{xt2}
\fmfshift{4 down}{xt2}
\fmfshift{4 right}{xt1}
\fmfshift{4 down}{xt1}
\fmfshift{4 left}{xb2}
\fmfshift{4 down}{xb2}
\fmfshift{4 right}{yb1}
\fmfshift{4 up}{yb1}
\fmfshift{4 left}{yt2}
\fmfshift{4 up}{yt2}
\fmfshift{4 right}{yt1}
\fmfshift{4 up}{yt1}
\fmfshift{4 left}{yb2}
\fmfshift{4 up}{yb2}
\fmfshift{4 left}{clb}
\fmfshift{4 down}{clb}
\fmfshift{4 left}{clt}
\fmfshift{4 up}{clt}
\fmfshift{4 right}{crb}
\fmfshift{4 down}{crb}
\fmfshift{4 right}{crt}
\fmfshift{4 up}{crt}
\fmfv{d.sh=circle,d.f=empty,d.si=11,background=(1,,.51,,.5)}{vb1}
\fmfv{d.sh=circle,d.f=empty,d.si=11,background=(.6235,,.7412,,1)}{vb2}
\fmfv{d.sh=circle,d.f=empty,d.si=11,background=(1,,.51,,.5)}{vt1}
\fmfv{d.sh=circle,d.f=empty,d.si=11,background=(.6235,,.7412,,1)}{vt2}
\fmfv{d.sh=circle,d.f=empty,d.si=11,background=(1,,.51,,.5)}{c}
\fmf{dashes}{l2,vb1}
\fmf{dashes}{l6,vt1}
\fmf{dbl_plain}{r2,vb2}
\fmf{dbl_plain}{r6,vt2}
\fmfset{arrow_len}{2.5mm}
\fmfi{fermion}{ vloc(__xt2).. vloc(__crt)}
\fmfi{fermion}{vloc(__xt1) ..vloc(__clt)}
\fmfi{fermion}{vloc(__crb).. vloc(__yb2)}
\fmfi{fermion}{vloc(__clb).. vloc(__yb1)}
\fmfi{fermion}{vloc(__yt2) .. vloc(__yt1)}
\fmfi{fermion}{vloc(__xb1) .. vloc(__xb2)}
\end{fmfgraph*}}
\nn[7mm]
&(d) \hspace{4mm}
\parbox{35mm}{
\begin{fmfgraph*}(30,15)
\fmfstraight
\fmfleftn{l}{7}\fmfrightn{r}{7}\fmfbottomn{b}{7}\fmftopn{t}{7}
\fmf{phantom}{l2,vb1,mb1,vb2,r2}
\fmf{phantom}{l6,vt1,mt1,vt2,r6}
\fmf{phantom}{l2,xb1,mb1,xb2,r2}
\fmf{phantom}{l6,xt1,mt1,xt2,r6}
\fmf{phantom}{l2,yb1,mb1,yb2,r2}
\fmf{phantom}{l6,yt1,mt1,yt2,r6}
\fmf{phantom}{l4,c,r4}
\fmf{phantom}{l4,clb,r4}
\fmf{phantom}{l4,clt,r4}
\fmf{phantom}{l4,crb,r4}
\fmf{phantom}{l4,crt,r4}
\fmffreeze
\fmfshift{4 right}{xb1}
\fmfshift{4 down}{xb1}
\fmfshift{4 left}{xt2}
\fmfshift{4 down}{xt2}
\fmfshift{4 right}{xt1}
\fmfshift{4 down}{xt1}
\fmfshift{4 left}{xb2}
\fmfshift{4 down}{xb2}
\fmfshift{4 right}{yb1}
\fmfshift{4 up}{yb1}
\fmfshift{4 left}{yt2}
\fmfshift{4 up}{yt2}
\fmfshift{4 right}{yt1}
\fmfshift{4 up}{yt1}
\fmfshift{4 left}{yb2}
\fmfshift{4 up}{yb2}
\fmfshift{4 left}{clb}
\fmfshift{4 down}{clb}
\fmfshift{4 left}{clt}
\fmfshift{4 up}{clt}
\fmfshift{4 right}{crb}
\fmfshift{4 down}{crb}
\fmfshift{4 right}{crt}
\fmfshift{4 up}{crt}
\fmfv{d.sh=circle,d.f=empty,d.si=11,background=(.6235,,.7412,,1)}{vb1}
\fmfv{d.sh=circle,d.f=empty,d.si=11,background=(1,,.51,,.5)}{vb2}
\fmfv{d.sh=circle,d.f=empty,d.si=11,background=(.6235,,.7412,,1)}{vt1}
\fmfv{d.sh=circle,d.f=empty,d.si=11,background=(1,,.51,,.5)}{vt2}
\fmfv{d.sh=circle,d.f=empty,d.si=11,background=(1,,.51,,.5)}{c}
\fmf{dbl_plain}{l2,vb1}
\fmf{dbl_plain}{l6,vt1}
\fmf{dashes}{r2,vb2}
\fmf{dashes}{r6,vt2}
\fmfset{arrow_len}{2.5mm}
\fmfi{fermion}{ vloc(__crt).. vloc(__xt2)}
\fmfi{fermion}{vloc(__clt) ..vloc(__xt1)}
\fmfi{fermion}{vloc(__yb2)..vloc(__crb) }
\fmfi{fermion}{ vloc(__yb1)..vloc(__clb)}
\fmfi{fermion}{vloc(__yt2) .. vloc(__yt1)}
\fmfi{fermion}{vloc(__xb1) .. vloc(__xb2)}
\end{fmfgraph*}}
\hspace{6mm}(e) \hspace{4mm}
\parbox{30mm}{
\begin{fmfgraph*}(30,15)
\fmfstraight
\fmfleftn{l}{7}\fmfrightn{r}{7}\fmfbottomn{b}{7}\fmftopn{t}{7}
\fmf{phantom}{l2,vb1,mb1,vb2,r2}
\fmf{phantom}{l6,vt1,mt1,vt2,r6}
\fmf{phantom}{l2,xb1,mb1,xb2,r2}
\fmf{phantom}{l6,xt1,mt1,xt2,r6}
\fmf{phantom}{l2,yb1,mb1,yb2,r2}
\fmf{phantom}{l6,yt1,mt1,yt2,r6}
\fmf{phantom}{l4,c,r4}
\fmf{phantom}{l4,clb,r4}
\fmf{phantom}{l4,clt,r4}
\fmf{phantom}{l4,crb,r4}
\fmf{phantom}{l4,crt,r4}
\fmffreeze
\fmfshift{4 right}{xb1}
\fmfshift{4 down}{xb1}
\fmfshift{4 left}{xt2}
\fmfshift{4 down}{xt2}
\fmfshift{4 right}{xt1}
\fmfshift{4 down}{xt1}
\fmfshift{4 left}{xb2}
\fmfshift{4 down}{xb2}
\fmfshift{4 right}{yb1}
\fmfshift{4 up}{yb1}
\fmfshift{4 left}{yt2}
\fmfshift{4 up}{yt2}
\fmfshift{4 right}{yt1}
\fmfshift{4 up}{yt1}
\fmfshift{4 left}{yb2}
\fmfshift{4 up}{yb2}
\fmfshift{4 left}{clb}
\fmfshift{4 down}{clb}
\fmfshift{4 left}{clt}
\fmfshift{4 up}{clt}
\fmfshift{4 right}{crb}
\fmfshift{4 down}{crb}
\fmfshift{4 right}{crt}
\fmfshift{4 up}{crt}
\fmfv{d.sh=circle,d.f=empty,d.si=11,background=(.6235,,.7412,,1)}{vb1}
\fmfv{d.sh=circle,d.f=empty,d.si=11,background=(.6235,,.7412,,1)}{vb2}
\fmfv{d.sh=circle,d.f=empty,d.si=11,background=(.6235,,.7412,,1)}{vt1}
\fmfv{d.sh=circle,d.f=empty,d.si=11,background=(.6235,,.7412,,1)}{vt2}
\fmfv{d.sh=circle,d.f=empty,d.si=11,background=(1,,.51,,.5)}{c}
\fmf{dbl_plain}{l2,vb1}
\fmf{dbl_plain}{l6,vt1}
\fmf{dbl_plain}{r2,vb2}
\fmf{dbl_plain}{r6,vt2}
\fmfset{arrow_len}{2.5mm}
\fmfi{fermion}{vloc(__xt2) .. vloc(__crt)}
\fmfi{fermion}{ vloc(__clt)..vloc(__xt1)}
\fmfi{fermion}{vloc(__crb) ..vloc(__yb2)}
\fmfi{fermion}{vloc(__yb1) ..vloc(__clb)}
\fmfi{fermion}{vloc(__yt2) .. vloc(__yt1)}
\fmfi{fermion}{vloc(__xb1) .. vloc(__xb2)}
\end{fmfgraph*}}
\end{align*}
\end{fmffile}   
\vspace{-3mm}

\caption{\fign{pots1}  Feynman diagrams making up the elements of the coupled channel $MM-D\bar D$ kernel matrix $V^{(1)}$ of \eq{V1mat}: (a) $\bGamma_1\PP_{12}T^+_1\PP_{12}\Gamma_1$, (b) $\bGamma_1\PP_{12}T^+_3\PP_{12}\Gamma_1$,  (c) $\bGamma_1\PP_{12}T^+_1 \Gamma_3$, (d) $\bGamma_3 T^+_1\PP_{12}\Gamma_1$, and (e) $\bGamma_3T^+_1\Gamma_3$.  Solid lines with leftward (rightward) arrows represent quarks (antiquarks), dashed lines represent mesons, and double-lines represent diquarks and antidiquarks.}
\end{center}
\end{figure}

\subsection{Meson-meson symmetry}
To discuss the symmetry of identical meson legs, we note that the potential $V$ consists of diagrams, some of which are illustrated in \fig{pots0} and \fig{pots1}, where a four-meson leg contribution, for example  $\bar\Gamma_1 {\cal P}_{12}\Gamma_1$ as illustrated in \fig{pots0}(a), consists of a diagram which is not symmetric with respect to meson quantum numbers, being only symmetric with respect to swapping meson legs  in both initial and final states simultaneously. Thus, to establish a description in terms of physical amplitudes, we will need to explicitly symmetrise identical meson states in the bound state equation,   \eq{400-}. To do this, we define ${\cal P}$ to be the operator that swaps meson quantum numbers, and note the useful relations
\begin{subequations} \eqn{useful}
\begin{align}
&{\cal P} \bGamma_1 = \bGamma_1{\cal P}_{12}{\cal P}_{34}, \\
&{\cal P}_{12}{\cal P}_{34} \Gamma_3 =\Gamma_3,
\end{align}
\end{subequations}
the first of which shows that interchanging the two mesons in the final state of the vertex function product $\bGamma_1 = \bGamma_{13}\bGamma_{24}$ is equivalent to interchanging the identical quarks and antiquarks in the initial state, and the second of which follows from the antisymmetry of the $qq$ and $\q\q$ vertex functions in $\Gamma_3 = \bGamma_{12}\bGamma_{34}$. Using these relations it is straightforward to prove 
\be
\begin{pmatrix} {\cal P} & 0\\  0& 1 \end{pmatrix} V =V \begin{pmatrix} {\cal P} & 0\\  0&1 \end{pmatrix}
\eqn{PV=VP}
\ee
which shows the equivalence of exchanging identical mesons in initial and final states. In turn this implies that if $\phi$ is a solution of   \eq{400-} then so is $\begin{pmatrix} {\cal P} & 0\\  0& 1 \end{pmatrix}\phi$, and therefore, so is
\be
\phi^S = \begin{pmatrix} 1+{\cal P} & 0 \\  0 &2 \end{pmatrix}\phi,
\ee
where $\phi^S$ is the physical solution which is symmetric with respect to the exchange of the two identical mesons. One can then write
\begin{align}
\phi^S &= V^S D  \phi^S  \eqn{BSS}
\end{align}
where
\be
V^S = \frac{1}{2}\begin{pmatrix} 1+{\cal P} & 0 \\  0 &2 \end{pmatrix} V
\ee
is the properly symmetrised kernel.  In particular,
\be
V^{S} = V^{S\, (0)} + V^{S\, (1)}     \eqn{VSum}
\ee
where
\begin{subequations} \eqn{VS}
\begin{align}
{V^{S\, (0)}}   &= - \begin{pmatrix} 1+{\cal P} & 0 \\  0 &2 \end{pmatrix} 
  \begin{pmatrix} -  \bar\Gamma_1{\cal P}_{12}  \Gamma_1& \bar\Gamma_1 \Gamma_3\\  
  \bar\Gamma_3  \Gamma_1&0 \end{pmatrix}  \eqn{Germ-her}, \\[3mm]
{V^{S\,(1)}}   &= -  \begin{pmatrix} 1+{\cal P} & 0 \\  0 &2 \end{pmatrix} \begin{pmatrix} \bar\Gamma_1[{\cal P}_{12}  T^+_1 {\cal P}_{12}+2T^+_3]\Gamma_1 &-\bar\Gamma_1{\cal P}_{12}  T^+_1\Gamma_3 \\ 
   -\bar\Gamma_3T^+_1{\cal P}_{12}\Gamma_1 & \bar\Gamma_3T^+_1 \Gamma_3\end{pmatrix} .
\end{align}
\end{subequations}
{According to the  discussion below \eqs{T3=T12},  $T_{12}=\frac{1}{2} T_{qq}$ and $T_{34}=\frac{1}{2} T_{\q\q}$, so that $T_3 = \frac{1}{2}(T_{qq} + T_{\q\q})$, where $T_{qq}$ and $T_{\q\q}$ are the physical (antisymmetric) scattering amplitudes for identical quarks. In the separable approximation {$ T_{qq}\sim i\Gamma_{qq}\bar\Gamma_{qq}/(P^2-M_{qq}^2)$ and $ T_{\q\q}\sim i\Gamma_{\q\q}\bar\Gamma_{\q\q}/(P^2-M_{\q\q}^2)$} which define the corresponding symmetrised quark vertex functions $\Gamma_{qq}$, $\Gamma_{\q\q}$, $\bGamma_{qq}$, and $\bGamma_{\q\q}$. It follows that $\Gamma_3=\frac{1}{2} \Gamma_{qq}\Gamma_{\bar q\bar q}$. It is convenient to re-express the symmetric (in mesons) kernels of \eqs{VS} in terms of these antisymmetric (in quarks) quantities. To do this in a way that does not change notation, we shall implement the following replacements:  $T_{12}\rightarrow \frac{1}{2} T_{12}$, $T_{34}\rightarrow \frac{1}{2} T_{34}$ and $\Gamma_3\rightarrow\frac{1}{2} \Gamma_3$. {After these replacements $T_{12}$ and $T_{34}$ become the physical scattering amplitudes for indistinguishable quarks and antiquarks.} In this way \eqs{VS} become
\begin{subequations} \eqn{VSa}
\begin{align}
{V^{S\, (0)}}   &= - \begin{pmatrix} 1+{\cal P} & 0 \\  0 &2 \end{pmatrix} 
  \begin{pmatrix} -  \bar\Gamma_1{\cal P}_{12}  \Gamma_1& \frac{1}{2} \bar\Gamma_1 \Gamma_3\\  
 \frac{1}{2}  \bar\Gamma_3  \Gamma_1&0 \end{pmatrix}, \\[3mm]
{V^{S\,(1)}}   &= -  \begin{pmatrix} 1+{\cal P} & 0 \\  0 &2 \end{pmatrix} \begin{pmatrix} \bar\Gamma_1[{\cal P}_{12}  T^+_1 {\cal P}_{12}+T^+_3]\Gamma_1 &-\frac{1}{2} \bar\Gamma_1{\cal P}_{12}  T^+_1\Gamma_3 \\ 
   -\frac{1}{2} \bar\Gamma_3T^+_1{\cal P}_{12}\Gamma_1 & \frac{1}{4} \bar\Gamma_3T^+_1 \Gamma_3\end{pmatrix} .
\end{align}
\end{subequations}
which using \eqs{useful}, simplify to
\begin{subequations} \eqn{VSa}
\begin{align}
{V^{S\, (0)}}   &= 
  \begin{pmatrix} (1+{\cal P} )\bar\Gamma_1{\cal P}_{12}  \Gamma_1&  - \bar\Gamma_1 \Gamma_3\\  
-  \bar\Gamma_3  \Gamma_1&0 \end{pmatrix},  \eqn{VS0} \\[3mm]
{V^{S\,(1)}}   &=  \begin{pmatrix} - (1+{\cal P} )\bar\Gamma_1[{\cal P}_{12}  T^+_1 {\cal P}_{12}+T^+_3]\Gamma_1 &\bar\Gamma_1{\cal P}_{12}  T^+_1\Gamma_3 \\ 
   \bar\Gamma_3T^+_1{\cal P}_{12}\Gamma_1 & -\frac{1}{2} \bar\Gamma_3T^+_1 \Gamma_3\end{pmatrix} . \eqn{VS1}
\end{align}
\end{subequations}
A few observations are in order:
\begin{enumerate}
\item {The expression for the lowest order potential,  $V^{S\,(0)}$, corresponds to the model of the Giessen group as previously derived  in \cite{Heupel:2012ua}. }

\item
One can see explicitly that the Giessen group potential $V^{S\,(0)}$ does not support $D\bar D$ elastic transition, $ D\bar D \leftarrow D\bar D$, whereas  the one of the Moscow group, $V^{S,(1)}$, does (see the right lower corner matrix element $2\Gamma_3T^+_1\Gamma_3$).

\item 
 Equation (\ref{VS1}) can be simplified by removing $T_{24}$ in $T^+_1 =T_{13}+T_{24}$, as follows. Using \eq{useful},
\begin{subequations}
\begin{align}
\bar\Gamma_1{\cal P}_{12}  T^+_1 {\cal P}_{12}\Gamma_1
&=\bar\Gamma_1{\cal P}_{12} ( T_{13}+T_{24}) {\cal P}_{12}\Gamma_1 \nn
&=\bar\Gamma_1{\cal P}_{12} ( T_{13}+{\cal P}_{12}{\cal P}_{34}T_{13}{\cal P}_{12}{\cal P}_{34}) {\cal P}_{12}\Gamma_1\nn
&  =\bar\Gamma_1{\cal P}_{12}  T_{13} {\cal P}_{12}\Gamma_1+{\cal P}\bar\Gamma_1{\cal P}_{12}  T_{13} {\cal P}_{12}\Gamma_1 {\cal P} 
\eqn{simp}\\[2mm]
\bGamma_1{\cal P}_{12}  T^+_1\Gamma_3&=\bGamma_1{\cal P}_{12}  T_{13}\Gamma_3+
{\cal P}\bar\Gamma_1 {\cal P}_{12}{\cal P}_{34}{\cal P}_{12} T_{24}\Gamma_3 \nn
& =\Gamma_1{\cal P}_{12}  T_{13}\Gamma_3+
{\cal P} \bar\Gamma_1 {\cal P}_{12} T_{13} \Gamma_3\nn
&=(1+{\cal P}) \bar\Gamma_1 {\cal P}_{12} T_{13} \Gamma_3,  \eqn{K+PKP}\\[2mm]
 \bar\Gamma_3  T^+_1 \Gamma_3 &= \bar\Gamma_3  (T_{13} + {\cal P}_{12}{\cal P}_{34}T_{24}{\cal P}_{12}{\cal P}_{34}) \Gamma_3 \nn
 & = 2 \bar\Gamma_3 T_{13}\Gamma_3.
\end{align}
\end{subequations}
The simplification is in that when solving numerically \eq{BSS}, instead of calculating two integrals of \eq{simp}, 
$\bar\Gamma_1{\cal P}_{12}  T_{13} {\cal P}_{12}\Gamma_1+\bar\Gamma_1{\cal P}_{34}  T_{13} {\cal P}_{34}\Gamma_1$, we calculate only one of them,  $I=\bar\Gamma_1{\cal P}_{12}  T_{13} {\cal P}_{12}\Gamma_1$, the second integral being obtained by only swapping meson quantum numbers in the first one, ${\cal P} I {\cal P}$. Similarly for $\bGamma_1{\cal P}_{12}  T^+_{1}\Gamma_3$.
\end{enumerate}

\section{Summary and discussion}
We have derived tetraquark equations that take the form of a Bethe-Salpeter equation in coupled $MM-D\bar D$ space, \eq{BSS}, where the kernel $V^S$ is a sum of two terms: $ V^{S\, (0)}$ consisting of terms involving non-interacting quark exchange, as
illustrated in \fig{pots0}, and  $ V^{S\, (1)}$ consisting of terms involving interacting quark exchange where one pair of quarks mutually scatter in intermediate state, as illustrated in \fig{pots1}. The mathematical expressions for these potentials are given by \eq{VSa}, which takes into account the antisymmetry of identical quarks ($qq$ and $\q\q$),  and the symmetry of identical mesons $(MM)$. 

Assuming pairwise interactions between the quarks, our derivation stems from the covariant four-body equations of Khvedelidze and Kvinikhidze \cite{Khvedelidze:1991qb}, which in this approximation, are exact equations for a four-body system in relativistic quantum field theory. Only two additional approximations are made to obtain our final equations: (i) separable approximations were made for each of the two-body t matrices in the product terms  $T_aT_{a'}$, of \eq{Ta}, thereby exposing $MM$ and $D\bar D$ channels, and (ii) the two-body t matrices in the sum  $T_a + T_{a'}$, of \eq{Ta}, are retained only to first order in the expression for the four-body kernel $V$, \eq{V}, which is sufficient to introduce  $q\bar q (T_{q\q})$, $q q (T_{\q\q})$, and $\q\q (T_{qq})$ states, as illustrated in \fig{pots1}, into the resulting description.

A feature of our equations, is that they provide a unified description of previous seemingly unrelated approaches. In particular, neglecting $V^{S\, (1)}$ from our kernel of \eq{VSum}, results in the $MM-D\bar D$ coupled channels model of the Giessen group (Fischer  {\it et al.})\! \cite{Heupel:2012ua,Eichmann:2015cra,Eichmann:2020oqt, Santowsky:2021bhy}, while neglecting $V^{S\, (0)}$ from our kernel of \eq{VSum}, encompasses the $D\bar D$ model of the Moscow group (Faustov {\it et al.})\! \cite{Ebert:2005nc,Faustov:2020qfm,Faustov:2021hjs,Faustov:2022mvs}. 
More specifically, the Moscow group model corresponds to keeping just the $D\D\rightarrow D\bar D$ element of the matrix $ V^{S\, (1)}$ given in \eq{VS1}, namely
\begin{align}
-\frac{1}{2}\bGamma_3 &T_1^+ \Gamma_3 = -\bGamma_{12}\bGamma_{34}  T_{13} \Gamma_{12}\Gamma_{34} \nn
&= -\bGamma_{D}\bGamma_{\bar D} G_{q\q}^0 T_{q\q} G_{q\q}^0  \Gamma_{D}\Gamma_{\bar D}    \eqn{DDkernel}
\end{align}
where $\Gamma_D\equiv \Gamma_{12}$, $\Gamma_{\bar D}\equiv \Gamma_{34}$, $\bGamma_D\equiv \bGamma_{12}$, $\bGamma_{\bar D}\equiv \bGamma_{34}$, $T_{q\q}\equiv T_{13}$, and  $G_{q\q}^0$ is the product of propagators for $q$ and $\q$. In this respect it is interesting to note that theory specifies $T_{q\q}$ to be the full t matrix for quark-antiquark scattering, and as such, is expressible as a sum of three types of contributions: (i) $s$-channel pole contributions corresponding to the formation of mesons {(the typical approximation used for two-quark scattering amplitudes by the Giessen group), (ii) a long-range contribution due to one-gluon-exchange}, and (iii) all other possible contribution including contributions responsible for confinement. Indeed, as shown in the Appendix, one can write the general structure of $T_{q\q}$ as
\be
T_{q\q} = \frac{\Phi_{q\q}\bar\Phi_{q\q}}{P^2-M_{q\q}^2}+ K_g + K_C 
\ee
where the pole term corresponds to a meson of mass $M_{q\q}$, $K_g$ is the one-gluon-exchange potential, and $K_C$ includes all other contributions {to $T_{q\q}$} including those responsible for confinement. Correspondingly, the $D\bar D$ kernel in our approach is given by the sum of the three terms illustrated in \fig{tqq}.
\begin{figure}[t]
\begin{center}
\begin{fmffile}{tqq}
\begin{align*}
&
\parbox{30mm}{
\begin{fmfgraph*}(30,15)
\fmfstraight
\fmfleftn{l}{7}\fmfrightn{r}{7}\fmfbottomn{b}{7}\fmftopn{t}{7}
\fmf{phantom}{l2,vb1,mb1,vb2,r2}
\fmf{phantom}{l6,vt1,mt1,vt2,r6}
\fmf{phantom}{l2,xb1,mb1,xb2,r2}
\fmf{phantom}{l6,xt1,mt1,xt2,r6}
\fmf{phantom}{l2,yb1,mb1,yb2,r2}
\fmf{phantom}{l6,yt1,mt1,yt2,r6}
\fmf{phantom}{l4,c,r4}
\fmf{phantom}{l4,clb,r4}
\fmf{phantom}{l4,clt,r4}
\fmf{phantom}{l4,crb,r4}
\fmf{phantom}{l4,crt,r4}
\fmffreeze
\fmfshift{4 right}{xb1}
\fmfshift{4 down}{xb1}
\fmfshift{4 left}{xt2}
\fmfshift{4 down}{xt2}
\fmfshift{4 right}{xt1}
\fmfshift{4 down}{xt1}
\fmfshift{4 left}{xb2}
\fmfshift{4 down}{xb2}
\fmfshift{4 right}{yb1}
\fmfshift{4 up}{yb1}
\fmfshift{4 left}{yt2}
\fmfshift{4 up}{yt2}
\fmfshift{4 right}{yt1}
\fmfshift{4 up}{yt1}
\fmfshift{4 left}{yb2}
\fmfshift{4 up}{yb2}
\fmfshift{4 left}{clb}
\fmfshift{4 down}{clb}
\fmfshift{4 left}{clt}
\fmfshift{4 up}{clt}
\fmfshift{4 right}{crb}
\fmfshift{4 down}{crb}
\fmfshift{4 right}{crt}
\fmfshift{4 up}{crt}
\fmfv{d.sh=circle,d.f=empty,d.si=11,background=(.6235,,.7412,,1)}{vb1}
\fmfv{d.sh=circle,d.f=empty,d.si=11,background=(.6235,,.7412,,1)}{vb2}
\fmfv{d.sh=circle,d.f=empty,d.si=11,background=(.6235,,.7412,,1)}{vt1}
\fmfv{d.sh=circle,d.f=empty,d.si=11,background=(.6235,,.7412,,1)}{vt2}
\fmfv{d.sh=circle,d.f=empty,d.si=11,background=(1,,.51,,.5)}{c}
\fmf{dbl_plain}{l2,vb1}
\fmf{dbl_plain}{l6,vt1}
\fmf{dbl_plain}{r2,vb2}
\fmf{dbl_plain}{r6,vt2}
\fmfset{arrow_len}{2.5mm}
\fmfi{fermion}{vloc(__xt2) .. vloc(__crt)}
\fmfi{fermion}{ vloc(__clt)..vloc(__xt1)}
\fmfi{fermion}{vloc(__crb) ..vloc(__yb2)}
\fmfi{fermion}{vloc(__yb1) ..vloc(__clb)}
\fmfi{fermion}{vloc(__yt2) .. vloc(__yt1)}
\fmfi{fermion}{vloc(__xb1) .. vloc(__xb2)}
\end{fmfgraph*}}
\hspace{2mm} = \hspace{2mm} 
\parbox{40mm}{
\begin{fmfgraph*}(40,15)
\fmfstraight
\fmfleftn{l}{7}\fmfrightn{r}{7}\fmfbottomn{b}{7}\fmftopn{t}{7}
\fmf{phantom}{l2,vb1}\fmf{phantom,tension=.25}{vb1,vb2}\fmf{phantom}{vb2,r2}
\fmf{phantom}{l2,ub1}\fmf{phantom,tension=.25}{ub1,ub2}\fmf{phantom}{ub2,r2}
\fmf{phantom}{l2,db1}\fmf{phantom,tension=.25}{db1,db2}\fmf{phantom}{db2,r2}
\fmf{phantom}{l2,rb1}\fmf{phantom,tension=.25}{rb1,rb2}\fmf{phantom}{rb2,r2}
\fmf{phantom}{l6,vt1}\fmf{phantom,tension=.25}{vt1,vt2}\fmf{phantom}{vt2,r6}
\fmf{phantom}{l6,ut1}\fmf{phantom,tension=.25}{ut1,ut2}\fmf{phantom}{ut2,r6}
\fmf{phantom}{l6,dt1}\fmf{phantom,tension=.25}{dt1,dt2}\fmf{phantom}{dt2,r6}
\fmffreeze
\fmf{phantom}{vt1,vm1}\fmf{phantom}{vb1,vm1}\fmf{phantom}{vt2,vm2}\fmf{phantom}{vb2,vm2}
\fmf{phantom,tension=1.2}{vm1,vm2}
\fmf{phantom}{ut1,um1}\fmf{phantom}{ub1,um1}\fmf{phantom}{ut2,um2}\fmf{phantom}{ub2,um2}
\fmf{phantom,tension=1.2}{um1,um2}
\fmf{phantom}{dt1,dm1}\fmf{phantom}{db1,dm1}\fmf{phantom}{dt2,dm2}\fmf{phantom}{db2,dm2}
\fmf{phantom,tension=1.2}{dm1,dm2}
\fmffreeze
\fmf{phantom}{vt1,rm1}\fmf{phantom}{vb1,rm1}\fmf{phantom}{vt2,rm2}\fmf{phantom}{vb2,rm2}
\fmf{phantom,tension=1.2}{rm1,rm2}
\fmffreeze
\fmfshift{4 right}{ub1}\fmfshift{4 up}{ub1}\fmfshift{4 right}{db1}\fmfshift{4 down}{db1}
\fmfshift{4 right}{ut1}\fmfshift{4 up}{ut1}\fmfshift{4 right}{dt1}\fmfshift{4 down}{dt1}
\fmfshift{4 left}{ub2}\fmfshift{4 up}{ub2}\fmfshift{4 left}{db2}\fmfshift{4 down}{db2}
\fmfshift{4 left}{ut2}\fmfshift{4 up}{ut2}\fmfshift{4 left}{dt2}\fmfshift{4 down}{dt2}
\fmfshift{4 left}{um1}\fmfshift{4 up}{um1}\fmfshift{4 left}{dm1}\fmfshift{4 down}{dm1}
\fmfshift{4 right}{um2}\fmfshift{4 up}{um2}\fmfshift{4 right}{dm2}\fmfshift{4 down}{dm2}
\fmfshift{1.4 left}{rm1}\fmfshift{1.4 left}{rm2}
\fmffreeze
\fmfv{d.sh=circle,d.f=empty,d.si=11,background=(.6235,,.7412,,1)}{vb1}
\fmfv{d.sh=circle,d.f=empty,d.si=11,background=(.6235,,.7412,,1)}{vb2}
\fmfv{d.sh=circle,d.f=empty,d.si=11,background=(.6235,,.7412,,1)}{vt1}
\fmfv{d.sh=circle,d.f=empty,d.si=11,background=(.6235,,.7412,,1)}{vt2}
\fmfv{d.sh=circle,d.f=empty,d.si=11,background=(1,,.51,,.5)}{vm1}
\fmfv{d.sh=circle,d.f=empty,d.si=11,background=(1,,.51,,.5)}{vm2}
\fmffreeze
\fmf{dbl_plain}{l2,vb1}
\fmf{dbl_plain}{l6,vt1}
\fmf{dbl_plain}{r2,vb2}
\fmf{dbl_plain}{r6,vt2}
\fmf{dashes}{rm1,rm2}
\fmfset{arrow_len}{2.5mm}
\fmf{fermion}{ut2,ut1}\fmf{fermion}{db1,db2}
\fmf{fermion}{um1,dt1}\fmf{fermion}{ub1,dm1}
\fmf{fermion}{dt2,um2}\fmf{fermion}{dm2,ub2}
\end{fmfgraph*}}
\hspace{2mm}+ \hspace{2mm}
\parbox{25mm}{
\begin{fmfgraph*}(25,25)
\fmfstraight
\fmfleftn{l}{7}\fmfrightn{r}{7}\fmfbottomn{b}{7}\fmftopn{t}{7}
\fmf{phantom}{l2,vb1}\fmf{phantom,tension=.45}{vb1,vb2}\fmf{phantom}{vb2,r2}
\fmf{phantom}{l2,ub1}\fmf{phantom,tension=.45}{ub1,ub2}\fmf{phantom}{ub2,r2}
\fmf{phantom}{l2,db1}\fmf{phantom,tension=.45}{db1,db2}\fmf{phantom}{db2,r2}
\fmf{phantom}{l2,rb1}\fmf{phantom,tension=.45}{rb1,rb2}\fmf{phantom}{rb2,r2}
\fmf{phantom}{l6,vt1}\fmf{phantom,tension=.45}{vt1,vt2}\fmf{phantom}{vt2,r6}
\fmf{phantom}{l6,ut1}\fmf{phantom,tension=.45}{ut1,ut2}\fmf{phantom}{ut2,r6}
\fmf{phantom}{l6,dt1}\fmf{phantom,tension=.45}{dt1,dt2}\fmf{phantom}{dt2,r6}
\fmffreeze
\fmfshift{4 right}{ub1}\fmfshift{4 up}{ub1}\fmfshift{4 right}{db1}\fmfshift{4 down}{db1}
\fmfshift{4 right}{ut1}\fmfshift{4 up}{ut1}\fmfshift{4 right}{dt1}\fmfshift{4 down}{dt1}
\fmfshift{4 left}{ub2}\fmfshift{4 up}{ub2}\fmfshift{4 left}{db2}\fmfshift{4 down}{db2}
\fmfshift{4 left}{ut2}\fmfshift{4 up}{ut2}\fmfshift{4 left}{dt2}\fmfshift{4 down}{dt2}
\fmffreeze
\fmf{phantom}{dt1,mt}\fmf{phantom}{dt2,mt}\fmf{phantom}{ub1,mb}\fmf{phantom}{ub2,mb}
\fmfset{curly_len}{2.5mm}
\fmf{curly,tension=.4}{mb,mt}
\fmfv{d.sh=circle,d.f=empty,d.si=11,background=(.6235,,.7412,,1)}{vb1}
\fmfv{d.sh=circle,d.f=empty,d.si=11,background=(.6235,,.7412,,1)}{vb2}
\fmfv{d.sh=circle,d.f=empty,d.si=11,background=(.6235,,.7412,,1)}{vt1}
\fmfv{d.sh=circle,d.f=empty,d.si=11,background=(.6235,,.7412,,1)}{vt2}
\fmfv{d.sh=circle,d.f=full,d.si=3}{mb}\fmfv{d.sh=circle,d.f=full,d.si=3}{mt}
\fmffreeze
\fmf{dbl_plain}{l2,vb1}
\fmf{dbl_plain}{l6,vt1}
\fmf{dbl_plain}{r2,vb2}
\fmf{dbl_plain}{r6,vt2}
\fmfset{arrow_len}{2.5mm}
\fmf{fermion}{ut2,ut1}\fmf{fermion}{db1,db2}
\fmf{fermion}{mt,dt1}\fmf{fermion}{dt2,mt}
\fmf{fermion}{ub1,mb}\fmf{fermion}{mb,ub2}
\end{fmfgraph*}}
\hspace{2mm} + \hspace{2mm}
\parbox{30mm}{
\begin{fmfgraph*}(30,15)
\fmfstraight
\fmfleftn{l}{7}\fmfrightn{r}{7}\fmfbottomn{b}{7}\fmftopn{t}{7}
\fmf{phantom}{l2,vb1,mb1,vb2,r2}
\fmf{phantom}{l6,vt1,mt1,vt2,r6}
\fmf{phantom}{l2,xb1,mb1,xb2,r2}
\fmf{phantom}{l6,xt1,mt1,xt2,r6}
\fmf{phantom}{l2,yb1,mb1,yb2,r2}
\fmf{phantom}{l6,yt1,mt1,yt2,r6}
\fmf{phantom}{l4,c,r4}
\fmf{phantom}{l4,clb,r4}
\fmf{phantom}{l4,clt,r4}
\fmf{phantom}{l4,crb,r4}
\fmf{phantom}{l4,crt,r4}
\fmffreeze
\fmfshift{4 right}{xb1}
\fmfshift{4 down}{xb1}
\fmfshift{4 left}{xt2}
\fmfshift{4 down}{xt2}
\fmfshift{4 right}{xt1}
\fmfshift{4 down}{xt1}
\fmfshift{4 left}{xb2}
\fmfshift{4 down}{xb2}
\fmfshift{4 right}{yb1}
\fmfshift{4 up}{yb1}
\fmfshift{4 left}{yt2}
\fmfshift{4 up}{yt2}
\fmfshift{4 right}{yt1}
\fmfshift{4 up}{yt1}
\fmfshift{4 left}{yb2}
\fmfshift{4 up}{yb2}
\fmfshift{4 left}{clb}
\fmfshift{4 down}{clb}
\fmfshift{4 left}{clt}
\fmfshift{4 up}{clt}
\fmfshift{4 right}{crb}
\fmfshift{4 down}{crb}
\fmfshift{4 right}{crt}
\fmfshift{4 up}{crt}
\fmfv{d.sh=circle,d.f=empty,d.si=11,background=(.6235,,.7412,,1)}{vb1}
\fmfv{d.sh=circle,d.f=empty,d.si=11,background=(.6235,,.7412,,1)}{vb2}
\fmfv{d.sh=circle,d.f=empty,d.si=11,background=(.6235,,.7412,,1)}{vt1}
\fmfv{d.sh=circle,d.f=empty,d.si=11,background=(.6235,,.7412,,1)}{vt2}
\fmfv{d.sh=circle,d.f=shaded,d.si=11,background=(1,,.51,,.5)}{c}
\fmf{dbl_plain}{l2,vb1}
\fmf{dbl_plain}{l6,vt1}
\fmf{dbl_plain}{r2,vb2}
\fmf{dbl_plain}{r6,vt2}
\fmfset{arrow_len}{2.5mm}
\fmfi{fermion}{vloc(__xt2) .. vloc(__crt)}
\fmfi{fermion}{ vloc(__clt)..vloc(__xt1)}
\fmfi{fermion}{vloc(__crb) ..vloc(__yb2)}
\fmfi{fermion}{vloc(__yb1) ..vloc(__clb)}
\fmfi{fermion}{vloc(__yt2) .. vloc(__yt1)}
\fmfi{fermion}{vloc(__xb1) .. vloc(__xb2)}
\end{fmfgraph*}}
\end{align*}
\end{fmffile}   
\vspace{-3mm}

\caption{\fign{tqq}  General structure of the $D\bar D$ kernel in the unified tetraquark equations. Illustrated is the  $D\bar D$ kernel (left diagram where the red circle represents the full $q\q$ t matrix $T_{q\q}$ in intermediate state), expressed as a sum of three terms (from left to right): (i) a $q\q$ $s$-channel meson exchange (dashed line) contribution, (ii) a $q\q$ one-gluon-exchange (curly line) contribution, and (iii) all possible other contributions to intermediate state $q\q$ scattering (shaded circle).}
\end{center}
\end{figure}
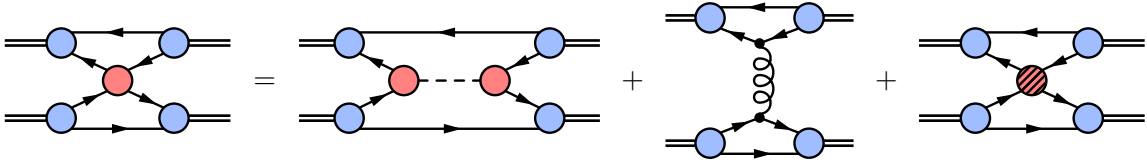

Comparison with the Moscow group's $D\bar D$ kernel shows that they did not include the $s$-channel meson exchange contribution (second diagram of \fig{tqq}), but did include one-gluon exchange taking into account the finite sizes of the diquark and antidiquark through corresponding form factors, [first term of Eq.\ (10) in Ref.\ \cite{Faustov:2021hjs}], a contribution corresponding to the third diagram of \fig{tqq}. The Moscow group also included a phenomenological $D\bar D $ confining potential  [second term of Eq.\ (10) in Ref.\ \cite{Faustov:2021hjs}], that correspond to the last diagram of \fig{tqq} for the case of a local $q\q$ potential.
Note that the confining interaction between a quark and an antiquark that are constitutents of a diquark and an antidiquark, results in diquark-antidiquark confinement, i.e.,   the two-body diquark-antidiquark potential produced in this way also has a confining part. Given the locality of the  $q\bar q$ confining potential, it only needs to be multiplied by a diquark form factor to result in the diquark-antidiquark confining potential, because the form factor does not change the long-range (small momentum transfer) behaviour of  the  $q\bar q$ potential.

Finally, it is worth noting that although we have singled out the works of the Moscow and Giessen groups as a means of demonstrating how our tetraquark equations can provide a common theoretical basis for very different approaches, it seems likely that these equations are able to encompass yet other theoretical works on the tetraquark.

\begin{acknowledgments}
 A.N.K. was supported by the Shota Rustaveli National Science Foundation (Grant No. FR17-354).

\end{acknowledgments}

\appendix

 \section{General structure of the $\bm{q\q}$ scattering amplitude}

Although it is not possible as yet to solve Quantum Chromodynamics to obtain the precise form of the force between a quark and an antiquark, there are three basic features of this force that would be desirable to take into account when constructing a phenomenological version of the $q\q$ scattering amplitude: (i) the force binds $q\q$ pairs to form mesons,  (ii) one-gluon-exchange is an important contribution to the {short-range} part of this force, and (iii) the force has the property of color confinement. To construct the $q\q$ t matrix $T$ with these features, one can first write the full $q\q$ Green function at total momentum $P$ in the form
\be
G =  \frac{\Psi \bPsi}{P^2 - M^2} +  G_C
\ee
where the pole term takes into account the bound state meson of mass $M$ (one can of course take into account more than one bound state by having a sum over such pole terms) and $G_C$ is the rest of the Green function with no pole at $P^2 = M^2$. If $K$ is the $q\q$ potential that generates $G$, that is if
\be
G = G_0 + G_0 K G,
\ee
then the corresponding t matrix $T$, defined as the solution of
\be
T = K+ K G_0 T,    \eqn{TK}
\ee
can be written as
\begin{align}
T& =K+KG K\nn[2mm]
& = K+K\left[\frac{\Psi\bar\Psi}{P^2-M^2}+G_C\right]K\nn[3mm]
& = K+\frac{\Phi\bar\Phi}{P^2-M^2}+KG_CK \eqn{T=K+KGK}
\end{align}
where $\Phi = K \Psi$. 
It is seen that the pole term is generated by the sum of the iterated terms of \eq{TK}, apart from $K$, i.e., the iteration series for the pole term starts with $KG_0K$. This means that adding the potential $K$ to the pole term does not overcount $K$, as one might otherwise expect. Writing $K$ as
\be
K = K_g + K_c
\ee
where $K_g$ is the one-gluon-exchange potential and $K_c\equiv K-K_g$, one  obtains the general structure of the $q\q$ t matrix:
\be
T = \frac{\Phi\bar\Phi}{P^2-M^2}+ K_g + K_C     \eqn{Tgen}
\ee
where
\be
K_C \equiv K_c + KG_CK
\ee
is responsible for confinement {in view of its contributions from $K_c$.}  As noted, neither $K_g$ nor $K_C$ is overcounted in \eq{Tgen}.

\bibliography{/Users/phbb/Physics/Papers/refn} 

\begin{thebibliography}{17}%
\makeatletter
\providecommand \@ifxundefined [1]{%
 \@ifx{#1\undefined}
}%
\providecommand \@ifnum [1]{%
 \ifnum #1\expandafter \@firstoftwo
 \else \expandafter \@secondoftwo
 \fi
}%
\providecommand \@ifx [1]{%
 \ifx #1\expandafter \@firstoftwo
 \else \expandafter \@secondoftwo
 \fi
}%
\providecommand \natexlab [1]{#1}%
\providecommand \enquote  [1]{``#1''}%
\providecommand \bibnamefont  [1]{#1}%
\providecommand \bibfnamefont [1]{#1}%
\providecommand \citenamefont [1]{#1}%
\providecommand \href@noop [0]{\@secondoftwo}%
\providecommand \href [0]{\begingroup \@sanitize@url \@href}%
\providecommand \@href[1]{\@@startlink{#1}\@@href}%
\providecommand \@@href[1]{\endgroup#1\@@endlink}%
\providecommand \@sanitize@url [0]{\catcode `\\12\catcode `\$12\catcode
  `\&12\catcode `\#12\catcode `\^12\catcode `\_12\catcode `\%12\relax}%
\providecommand \@@startlink[1]{}%
\providecommand \@@endlink[0]{}%
\providecommand \url  [0]{\begingroup\@sanitize@url \@url }%
\providecommand \@url [1]{\endgroup\@href {#1}{\urlprefix }}%
\providecommand \urlprefix  [0]{URL }%
\providecommand \Eprint [0]{\href }%
\providecommand \doibase [0]{https://doi.org/}%
\providecommand \selectlanguage [0]{\@gobble}%
\providecommand \bibinfo  [0]{\@secondoftwo}%
\providecommand \bibfield  [0]{\@secondoftwo}%
\providecommand \translation [1]{[#1]}%
\providecommand \BibitemOpen [0]{}%
\providecommand \bibitemStop [0]{}%
\providecommand \bibitemNoStop [0]{.\EOS\space}%
\providecommand \EOS [0]{\spacefactor3000\relax}%
\providecommand \BibitemShut  [1]{\csname bibitem#1\endcsname}%
\let\auto@bib@innerbib\@empty
\bibitem [{\citenamefont {Gell-Mann}(1964)}]{Gell-Mann:1964ewy}%
  \BibitemOpen
  \bibfield  {author} {\bibinfo {author} {\bibfnamefont {M.}~\bibnamefont
  {Gell-Mann}},\ }\bibfield  {title} {\bibinfo {title} {{A Schematic Model of
  Baryons and Mesons}},\ }\href {https://doi.org/10.1016/S0031-9163(64)92001-3}
  {\bibfield  {journal} {\bibinfo  {journal} {Phys. Lett.}\ }\textbf {\bibinfo
  {volume} {8}},\ \bibinfo {pages} {214} (\bibinfo {year} {1964})}\BibitemShut
  {NoStop}%
\bibitem [{\citenamefont {Zweig}(1964)}]{Zweig:1964ruk}%
  \BibitemOpen
  \bibfield  {author} {\bibinfo {author} {\bibfnamefont {G.}~\bibnamefont
  {Zweig}},\ }\bibfield  {title} {\bibinfo {title} {{An SU(3) model for strong
  interaction symmetry and its breaking. Version 1}},\ }\bibfield  {journal}
  {\bibinfo  {journal} {Report No. CERN-TH-401,}\ }\href
  {https://doi.org/https://cds.cern.ch/record/352337/files/CERN-TH-401.pdf}
  {https://cds.cern.ch/record/352337/files/CERN-TH-401.pdf} (\bibinfo {year}
  {1964})\BibitemShut {NoStop}%
\bibitem [{\citenamefont {Jaffe}(1977{\natexlab{a}})}]{Jaffe_PRD15_267}%
  \BibitemOpen
  \bibfield  {author} {\bibinfo {author} {\bibfnamefont {R.~J.}\ \bibnamefont
  {Jaffe}},\ }\bibfield  {title} {\bibinfo {title} {Multiquark hadrons. i.
  phenomenology of ${Q}^{2}{\overline{q}}^{2}$ mesons},\ }\href
  {https://doi.org/10.1103/PhysRevD.15.267} {\bibfield  {journal} {\bibinfo
  {journal} {Phys. Rev. D}\ }\textbf {\bibinfo {volume} {15}},\ \bibinfo
  {pages} {267} (\bibinfo {year} {1977}{\natexlab{a}})}\BibitemShut {NoStop}%
\bibitem [{\citenamefont {Jaffe}(1977{\natexlab{b}})}]{Jaffe_PRD15_281}%
  \BibitemOpen
  \bibfield  {author} {\bibinfo {author} {\bibfnamefont {R.~L.}\ \bibnamefont
  {Jaffe}},\ }\bibfield  {title} {\bibinfo {title} {Multiquark hadrons. ii.
  methods},\ }\href {https://doi.org/10.1103/PhysRevD.15.281} {\bibfield
  {journal} {\bibinfo  {journal} {Phys. Rev. D}\ }\textbf {\bibinfo {volume}
  {15}},\ \bibinfo {pages} {281} (\bibinfo {year}
  {1977}{\natexlab{b}})}\BibitemShut {NoStop}%
\bibitem [{\citenamefont {Choi}\ \emph {et~al.}(2003)\citenamefont {Choi} \emph
  {et~al.}}]{Belle:2003nnu}%
  \BibitemOpen
  \bibfield  {author} {\bibinfo {author} {\bibfnamefont {S.~K.}\ \bibnamefont
  {Choi}} \emph {et~al.} (\bibinfo {collaboration} {Belle}),\ }\bibfield
  {title} {\bibinfo {title} {{Observation of a narrow charmonium-like state in
  exclusive $B^\pm \to K^\pm \pi^+ \pi^- J/\psi$ decays}},\ }\href
  {https://doi.org/10.1103/PhysRevLett.91.262001} {\bibfield  {journal}
  {\bibinfo  {journal} {Phys. Rev. Lett.}\ }\textbf {\bibinfo {volume} {91}},\
  \bibinfo {pages} {262001} (\bibinfo {year} {2003})},\ \Eprint
  {https://arxiv.org/abs/hep-ex/0309032} {arXiv:hep-ex/0309032} \BibitemShut
  {NoStop}%
\bibitem [{\citenamefont {Chen}\ \emph {et~al.}(2023)\citenamefont {Chen},
  \citenamefont {Chen}, \citenamefont {Liu}, \citenamefont {Liu},\ and\
  \citenamefont {Zhu}}]{Chen:2022asf}%
  \BibitemOpen
  \bibfield  {author} {\bibinfo {author} {\bibfnamefont {H.-X.}\ \bibnamefont
  {Chen}}, \bibinfo {author} {\bibfnamefont {W.}~\bibnamefont {Chen}}, \bibinfo
  {author} {\bibfnamefont {X.}~\bibnamefont {Liu}}, \bibinfo {author}
  {\bibfnamefont {Y.-R.}\ \bibnamefont {Liu}},\ and\ \bibinfo {author}
  {\bibfnamefont {S.-L.}\ \bibnamefont {Zhu}},\ }\bibfield  {title} {\bibinfo
  {title} {{An updated review of the new hadron states}},\ }\href
  {https://doi.org/10.1088/1361-6633/aca3b6} {\bibfield  {journal} {\bibinfo
  {journal} {Rept. Prog. Phys.}\ }\textbf {\bibinfo {volume} {86}},\ \bibinfo
  {pages} {026201} (\bibinfo {year} {2023})},\ \Eprint
  {https://arxiv.org/abs/2204.02649} {arXiv:2204.02649 [hep-ph]} \BibitemShut
  {NoStop}%
\bibitem [{\citenamefont {Ebert}\ \emph {et~al.}(2006)\citenamefont {Ebert},
  \citenamefont {Faustov},\ and\ \citenamefont {Galkin}}]{Ebert:2005nc}%
  \BibitemOpen
  \bibfield  {author} {\bibinfo {author} {\bibfnamefont {D.}~\bibnamefont
  {Ebert}}, \bibinfo {author} {\bibfnamefont {R.~N.}\ \bibnamefont {Faustov}},\
  and\ \bibinfo {author} {\bibfnamefont {V.~O.}\ \bibnamefont {Galkin}},\
  }\bibfield  {title} {\bibinfo {title} {{Masses of heavy tetraquarks in the
  relativistic quark model}},\ }\href
  {https://doi.org/10.1016/j.physletb.2006.01.026} {\bibfield  {journal}
  {\bibinfo  {journal} {Phys. Lett. B}\ }\textbf {\bibinfo {volume} {634}},\
  \bibinfo {pages} {214} (\bibinfo {year} {2006})},\ \Eprint
  {https://arxiv.org/abs/hep-ph/0512230} {arXiv:hep-ph/0512230} \BibitemShut
  {NoStop}%
\bibitem [{\citenamefont {Faustov}\ \emph {et~al.}(2020)\citenamefont
  {Faustov}, \citenamefont {Galkin},\ and\ \citenamefont
  {Savchenko}}]{Faustov:2020qfm}%
  \BibitemOpen
  \bibfield  {author} {\bibinfo {author} {\bibfnamefont {R.~N.}\ \bibnamefont
  {Faustov}}, \bibinfo {author} {\bibfnamefont {V.~O.}\ \bibnamefont
  {Galkin}},\ and\ \bibinfo {author} {\bibfnamefont {E.~M.}\ \bibnamefont
  {Savchenko}},\ }\bibfield  {title} {\bibinfo {title} {{Masses of the $QQ\bar
  Q\bar Q$ tetraquarks in the relativistic diquark--antidiquark picture}},\
  }\href {https://doi.org/10.1103/PhysRevD.102.114030} {\bibfield  {journal}
  {\bibinfo  {journal} {Phys. Rev. D}\ }\textbf {\bibinfo {volume} {102}},\
  \bibinfo {pages} {114030} (\bibinfo {year} {2020})},\ \Eprint
  {https://arxiv.org/abs/2009.13237} {arXiv:2009.13237 [hep-ph]} \BibitemShut
  {NoStop}%
\bibitem [{\citenamefont {Faustov}\ \emph {et~al.}(2021)\citenamefont
  {Faustov}, \citenamefont {Galkin},\ and\ \citenamefont
  {Savchenko}}]{Faustov:2021hjs}%
  \BibitemOpen
  \bibfield  {author} {\bibinfo {author} {\bibfnamefont {R.~N.}\ \bibnamefont
  {Faustov}}, \bibinfo {author} {\bibfnamefont {V.~O.}\ \bibnamefont
  {Galkin}},\ and\ \bibinfo {author} {\bibfnamefont {E.~M.}\ \bibnamefont
  {Savchenko}},\ }\bibfield  {title} {\bibinfo {title} {{Heavy tetraquarks in
  the relativistic quark model}},\ }\href
  {https://doi.org/10.3390/universe7040094} {\bibfield  {journal} {\bibinfo
  {journal} {Universe}\ }\textbf {\bibinfo {volume} {7}},\ \bibinfo {pages}
  {94} (\bibinfo {year} {2021})},\ \Eprint {https://arxiv.org/abs/2103.01763}
  {arXiv:2103.01763 [hep-ph]} \BibitemShut {NoStop}%
\bibitem [{\citenamefont {Faustov}\ \emph {et~al.}(2022)\citenamefont
  {Faustov}, \citenamefont {Galkin},\ and\ \citenamefont
  {Savchenko}}]{Faustov:2022mvs}%
  \BibitemOpen
  \bibfield  {author} {\bibinfo {author} {\bibfnamefont {R.~N.}\ \bibnamefont
  {Faustov}}, \bibinfo {author} {\bibfnamefont {V.~O.}\ \bibnamefont
  {Galkin}},\ and\ \bibinfo {author} {\bibfnamefont {E.~M.}\ \bibnamefont
  {Savchenko}},\ }\bibfield  {title} {\bibinfo {title} {{Fully Heavy Tetraquark
  Spectroscopy in the Relativistic Quark Model}},\ }\href
  {https://doi.org/10.3390/sym14122504} {\bibfield  {journal} {\bibinfo
  {journal} {Symmetry}\ }\textbf {\bibinfo {volume} {14}},\ \bibinfo {pages}
  {2504} (\bibinfo {year} {2022})},\ \Eprint {https://arxiv.org/abs/2210.16015}
  {arXiv:2210.16015 [hep-ph]} \BibitemShut {NoStop}%
\bibitem [{\citenamefont {Heupel}\ \emph {et~al.}(2012)\citenamefont {Heupel},
  \citenamefont {Eichmann},\ and\ \citenamefont {Fischer}}]{Heupel:2012ua}%
  \BibitemOpen
  \bibfield  {author} {\bibinfo {author} {\bibfnamefont {W.}~\bibnamefont
  {Heupel}}, \bibinfo {author} {\bibfnamefont {G.}~\bibnamefont {Eichmann}},\
  and\ \bibinfo {author} {\bibfnamefont {C.~S.}\ \bibnamefont {Fischer}},\
  }\bibfield  {title} {\bibinfo {title} {{Tetraquark Bound States in a
  Bethe-Salpeter Approach}},\ }\href
  {https://doi.org/dx.doi.org/10.1016/j.physletb.2012.11.009} {\bibfield
  {journal} {\bibinfo  {journal} {Phys. Lett. B}\ }\textbf {\bibinfo {volume}
  {718}},\ \bibinfo {pages} {545} (\bibinfo {year} {2012})},\ \Eprint
  {https://arxiv.org/abs/1206.5129} {arXiv:1206.5129 [hep-ph]} \BibitemShut
  {NoStop}%
\bibitem [{\citenamefont {Eichmann}\ \emph {et~al.}(2016)\citenamefont
  {Eichmann}, \citenamefont {Fischer},\ and\ \citenamefont
  {Heupel}}]{Eichmann:2015cra}%
  \BibitemOpen
  \bibfield  {author} {\bibinfo {author} {\bibfnamefont {G.}~\bibnamefont
  {Eichmann}}, \bibinfo {author} {\bibfnamefont {C.~S.}\ \bibnamefont
  {Fischer}},\ and\ \bibinfo {author} {\bibfnamefont {W.}~\bibnamefont
  {Heupel}},\ }\bibfield  {title} {\bibinfo {title} {{The light scalar mesons
  as tetraquarks}},\ }\href {https://doi.org/10.1016/j.physletb.2015.12.036}
  {\bibfield  {journal} {\bibinfo  {journal} {Phys. Lett. B}\ }\textbf
  {\bibinfo {volume} {753}},\ \bibinfo {pages} {282} (\bibinfo {year}
  {2016})},\ \Eprint {https://arxiv.org/abs/1508.07178} {arXiv:1508.07178
  [hep-ph]} \BibitemShut {NoStop}%
\bibitem [{\citenamefont {Eichmann}\ \emph {et~al.}(2020)\citenamefont
  {Eichmann}, \citenamefont {Fischer}, \citenamefont {Heupel}, \citenamefont
  {Santowsky},\ and\ \citenamefont {Wallbott}}]{Eichmann:2020oqt}%
  \BibitemOpen
  \bibfield  {author} {\bibinfo {author} {\bibfnamefont {G.}~\bibnamefont
  {Eichmann}}, \bibinfo {author} {\bibfnamefont {C.~S.}\ \bibnamefont
  {Fischer}}, \bibinfo {author} {\bibfnamefont {W.}~\bibnamefont {Heupel}},
  \bibinfo {author} {\bibfnamefont {N.}~\bibnamefont {Santowsky}},\ and\
  \bibinfo {author} {\bibfnamefont {P.~C.}\ \bibnamefont {Wallbott}},\
  }\bibfield  {title} {\bibinfo {title} {{Four-Quark States from Functional
  Methods}},\ }\href {https://doi.org/10.1007/s00601-020-01571-3} {\bibfield
  {journal} {\bibinfo  {journal} {Few Body Syst.}\ }\textbf {\bibinfo {volume}
  {61}},\ \bibinfo {pages} {38} (\bibinfo {year} {2020})},\ \Eprint
  {https://arxiv.org/abs/2008.10240} {arXiv:2008.10240 [hep-ph]} \BibitemShut
  {NoStop}%
\bibitem [{\citenamefont {Santowsky}\ and\ \citenamefont
  {Fischer}(2022)}]{Santowsky:2021bhy}%
  \BibitemOpen
  \bibfield  {author} {\bibinfo {author} {\bibfnamefont {N.}~\bibnamefont
  {Santowsky}}\ and\ \bibinfo {author} {\bibfnamefont {C.~S.}\ \bibnamefont
  {Fischer}},\ }\bibfield  {title} {\bibinfo {title} {{Four-quark states with
  charm quarks in a two-body Bethe\textendash{}Salpeter approach}},\ }\href
  {https://doi.org/10.1140/epjc/s10052-022-10272-6} {\bibfield  {journal}
  {\bibinfo  {journal} {Eur. Phys. J. C}\ }\textbf {\bibinfo {volume} {82}},\
  \bibinfo {pages} {313} (\bibinfo {year} {2022})},\ \Eprint
  {https://arxiv.org/abs/2111.15310} {arXiv:2111.15310 [hep-ph]} \BibitemShut
  {NoStop}%
\bibitem [{\citenamefont {Khvedelidze}\ and\ \citenamefont
  {Kvinikhidze}(1992)}]{Khvedelidze:1991qb}%
  \BibitemOpen
  \bibfield  {author} {\bibinfo {author} {\bibfnamefont {A.~M.}\ \bibnamefont
  {Khvedelidze}}\ and\ \bibinfo {author} {\bibfnamefont {A.~N.}\ \bibnamefont
  {Kvinikhidze}},\ }\bibfield  {title} {\bibinfo {title} {Pair interaction
  approximation in the equations of quantum field theory for a four-body
  system},\ }\href@noop {} {\bibfield  {journal} {\bibinfo  {journal} {Theor.
  Math. Phys.}\ }\textbf {\bibinfo {volume} {90}},\ \bibinfo {pages} {62}
  (\bibinfo {year} {1992})}\BibitemShut {NoStop}%
\bibitem [{\citenamefont {Kvinikhidze}\ and\ \citenamefont
  {Blankleider}(2014)}]{Kvinikhidze:2014yqa}%
  \BibitemOpen
  \bibfield  {author} {\bibinfo {author} {\bibfnamefont {A.~N.}\ \bibnamefont
  {Kvinikhidze}}\ and\ \bibinfo {author} {\bibfnamefont {B.}~\bibnamefont
  {Blankleider}},\ }\bibfield  {title} {\bibinfo {title} {{Covariant equations
  for the tetraquark and more}},\ }\href
  {https://doi.org/10.1103/PhysRevD.90.045042} {\bibfield  {journal} {\bibinfo
  {journal} {Phys. Rev. D}\ }\textbf {\bibinfo {volume} {90}},\ \bibinfo
  {pages} {045042} (\bibinfo {year} {2014})},\ \Eprint
  {https://arxiv.org/abs/1406.5599} {arXiv:1406.5599 [hep-ph]} \BibitemShut
  {NoStop}%
\bibitem [{\citenamefont {Kvinikhidze}\ and\ \citenamefont
  {Blankleider}(2022)}]{Kvinikhidze:2021kzu}%
  \BibitemOpen
  \bibfield  {author} {\bibinfo {author} {\bibfnamefont {A.~N.}\ \bibnamefont
  {Kvinikhidze}}\ and\ \bibinfo {author} {\bibfnamefont {B.}~\bibnamefont
  {Blankleider}},\ }\bibfield  {title} {\bibinfo {title} {{Covariant tetraquark
  equations in quantum field theory}},\ }\href
  {https://doi.org/10.1103/PhysRevD.106.054024} {\bibfield  {journal} {\bibinfo
   {journal} {Phys. Rev. D}\ }\textbf {\bibinfo {volume} {106}},\ \bibinfo
  {pages} {054024} (\bibinfo {year} {2022})},\ \Eprint
  {https://arxiv.org/abs/2102.09558} {arXiv:2102.09558 [hep-th]} \BibitemShut
  {NoStop}%
\end{thebibliography}%

\end{document}